\documentclass[11pt,draftclsnofoot, onecolumn]{IEEEtran}

\usepackage{times} \usepackage{amssymb} \usepackage{amsmath}
\usepackage{epsfig} \usepackage{cite} \usepackage{verbatim}
\usepackage{subfigure}\usepackage{bm}

\psfull

\begin{document}

\renewcommand{\textfraction}{0} \newcommand{\albm}{{\bm{\alpha}}}
\newcommand{\Pbm}{{\bm{P}}} \newcommand{\Rbm}{{\bm{R}}}
\newcommand{\Pbmover}{{\overline{\Pbm}}} \newcommand{\hbm}{{\bm{h}}}
\newcommand{\Hbm}{{\bm{H}}}
\newcommand{\bh}{\bf{h}}
\newtheorem{theorem}{Theorem} \newtheorem{lemma}{Lemma}
\newtheorem{definition}{Definition} \newtheorem{corollary}{Corollary}

\title{MIMO Broadcast Channels with Finite Rate Feedback}
\author{\authorblockN{Nihar Jindal} \\
\authorblockA{Dept. of Electrical \& Computer Engineering \\
University of Minnesota\\
Minnesota, MN 55455 \\
Email: nihar@umn.edu, Phone: (612) 625-6306}}

 \maketitle

\thispagestyle{empty}


\begin{abstract}
Multiple transmit antennas in a downlink channel
can provide tremendous capacity (i.e. multiplexing) gains, even
when receivers have only single antennas.
However, receiver and transmitter channel state information
is generally required.  In this paper,
a system where each receiver has perfect channel knowledge,
but the transmitter only receives quantized information regarding the
channel instantiation is analyzed.  The well known zero forcing
transmission technique is considered, and simple expressions
for the throughput degradation due to finite rate feedback are
derived.  A key finding is that the feedback rate per mobile must be 
increased linearly with the SNR (in dB) in order to achieve
the full multiplexing gain, which is in sharp contrast to 
point-to-point MIMO systems in which it is not necessary to
increase the feedback rate as a function of the SNR.
\end{abstract}


\maketitle

\section{Introduction}

In multiple antenna broadcast (downlink) channels,  capacity can
be tremendously increased by adding antennas
at only the access point \cite{Caire_Shamai}\cite{Jindal_Goldsmith_DPC}.
In essence, an access point (AP) equipped with $M$ antennas
can support downlink rates up to a factor of $M$ times
larger than a single antenna access point, even when
each mobile device has only a single antenna\footnote{In fact,
this is true on the uplink as well, by the multiple-access/broadcast
channel duality \cite{Jindal_Goldsmith_DPC}}.  
In order to realize these benefits, however, the access point must:
\begin{itemize}
\item Simultaneously transmit to multiple users over the
same bandwidth (orthogonal schemes such as TDMA or CDMA
are generally highly sub-optimal).
\item Obtain accurate channel state information (CSI).
\end{itemize}
Practical transmission structures that allow for simultaneous
transmission to multiple mobiles, such as downlink beamforming,
do exist.  The requirement that the AP have
accurate CSI, however, is far more difficult to meet, particularly
in frequency division duplexed (FDD) systems.  Training can be used to 
obtain channel knowledge at
each of the mobile devices, but obtaining CSI at the AP
generally requires feedback from each mobile.
Such feedback channels do exist in current systems (e.g., for power
control), but the required rate of feedback is clearly an important
quantity for system designers.

In this paper, we consider the practically motivated
\textit{finite rate feedback} model, in which each mobile feeds back a finite
number of bits regarding its channel instantiation at the beginning
of each block or frame.
This model was first considered for
point-to-point MIMO channels in \cite{Narula_Trott1} \cite{Love_Heath}
\cite{Mukkavilli_MIMO}, where the transmitter uses such feedback to more
accurately direct its transmitted energy towards the receiver's antenna
array, and even a small number of bits per antenna
can be quite beneficial \cite{Love_Heath_Santipach_Honig}.
In point-to-point MIMO channels,
the level of CSI available at the transmitter only
affects the SNR-offset; it does not affect
the slope of the capacity vs. SNR curve, i.e., the
\textit{multiplexing gain}.
However, the level of CSI available to the transmitter
does affect the multiplexing gain of the MIMO downlink channel
\cite{Caire_Shamai}.  As a result,
channel feedback is considerably more
important for MIMO downlink channels
than for point-to-point channels.

In contrast to most recent work on the MIMO downlink channel
which has primarily concentrated on channels with a very large 
number of mobiles
\cite{Sharif_Hassibi}\cite{Yoo_Goldsmith}\cite{Swannack_Wornell},
we consider systems in which the number of
mobiles is equal to the number of transmit antennas.
This regime is applicable for inherently smaller systems, as well
as large systems in which stringent delay constraints
do not allow a small subset of users to be selected for transmission.
Random beamforming is an alternative limited feedback strategy
for MIMO downlink channels in which each mobile feeds back a
very low rate quantization of the channel ($\log_2 M$ bits)
as well as an analog SNR value \cite{Sharif_Hassibi}.
While this strategy performs well when there are a 
large number of mobiles relative to the number of transmit
antennas, it performs poorly in the small system regime which
we consider.

In this work we propose a simple downlink transmission scheme
that uses zero-forcing precoding in conjunction
with finite rate feedback.  We consider systems in which
each mobile performs vector quantization on its channel realization
using random quantization codebooks, i.e. \textit{random vector quantization}
\cite{Santipach_Honig_CDMA}\cite{Santipach_Honig}.
Our key findings are:
\begin{itemize}
\item The throughput of a feedback-based zero-forcing system is bounded
if the SNR is taken to infinity and the number of feedback bits per mobile is 
kept fixed.
\item The number of feedback bits per mobile ($B$) must be increased \textit{linearly}
with the SNR (in dB) at the rate
\begin{eqnarray*}
B = (M-1) \log_2 P 
&=& \frac{ (M-1) \log_2 10}{10} P_{dB} \\
&\approx& \frac{M-1}{3} P_{dB}
\end{eqnarray*}
in order to achieve the full multiplexing gain of $M$.   In addition,
this scaling of $B$ guarantees that
the throughput loss relative to perfect CSIT-based zero-forcing is
upper bounded by $M$ bps/Hz, which corresponds to a 3 dB power offset.
\item
Scaling the number of feedback bits according to
$B = \alpha \log_2 P$ for any $\alpha < M-1$ results in
a strictly inferior multiplexing
gain of $M\left(\frac{\alpha}{M-1}\right)$.
\end{itemize}
In essence, the channel estimation error at the access point must
scale as the inverse of the SNR in order to allow the full
multiplexing gain to be achieved, which results in the required
linear scaling of feedback.  As a result of this scaling,
feedback requirements are considerably higher in downlink channels than
in point-to-point MIMO channels, which will have considerable
design implications.

The MIMO downlink finite-rate feedback
model was also considered independently by Ding, Love, and Zoltowski
\cite{Ding_Love_Zoltowski}

The remainder of this paper is organized as follows.
In Section \ref{sec-system_model} we describe the channel model
and the finite rate feedback mechanism.  In 
Section \ref{sec-background} we provide background material
on MIMO downlink capacity, downlink precoding, and random
vector quantization.  In Section \ref{sec-sys_description}
we described the proposed zero-forcing based system, and 
analyze the throughput of this system in Section 
\ref{sec-throughput}.  We provide numerical results
comparing finite-rate feedback systems to alternative
transmission techniques in Section \ref{sec-comparison}, and
close by discussing conclusions and possible extensions of this work in
Section \ref{sec-conclusions}.

\section{System Model} \label{sec-system_model}

We consider a $K$ receiver multiple antenna broadcast
channel in which the transmitter (access point, or AP) has
$M > 1$ antennas and each receiver has a single antenna.
The broadcast channel is mathematically described as:
\begin{equation}
y_i = {\mathbf h}_i^{\dagger} {\mathbf x} + n_i, ~~~
   i=1,\ldots,K
\end{equation}
where ${\bf h}_1, {\bf h}_2, \ldots, {\bf
        h}_K$ are the channel vectors (with ${\bf h}_i \in
        {\mathbb{C}}^{M \times 1}$) of users 1 through $K$,
the vector ${\bf x} \in
        {\mathbb{C}}^{M \times 1}$  is the transmitted
        signal, and ${\bf n}_1, \ldots, {\mathbf n}_K$
are independent complex Gaussian noise terms
with unit variance.
The input must satisfy a transmit power constraint of $P$, i.e.
$E[||{\mathbf x}||^2] \leq P$.
We denote the concatenation of the channels by
${\mathbf H}^{\dagger} = [{\mathbf h}_1 ~ {\mathbf h}_2 \cdots {\mathbf h}_K]$,
i.e. ${\mathbf H}$ is $K \times M$ with the $i$-th row equal to
the channel of the $i$-th receiver (${\mathbf h}_i^{\dagger}$).
\begin{figure*}
\centering
\centerline{\epsfig{file=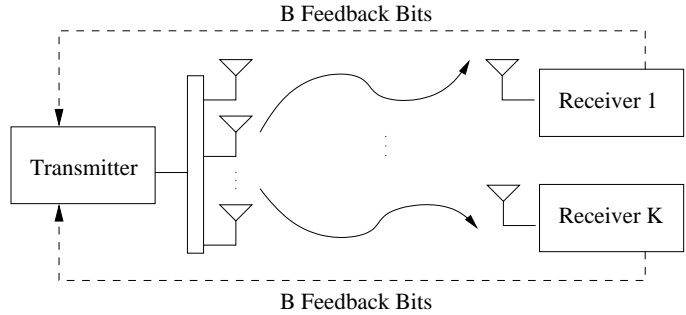, width=90mm}}
\caption{Finite Rate Feedback System Model}
\label{fig-model}
\end{figure*}
In order to focus our efforts on the impact of 
imperfect CSI, we consider a system where the number of
mobiles is equal to the number of transmit antennas, 
i.e., $K=M$.  
However, this work can be combined
with so-called user selection algorithms for systems with
more mobiles than antennas ($K > M$), as studied in
\cite{Yoo_Jindal_Goldsmith}.

The channel is assumed to be block fading, with
independent fading from block to block.  The entries of
the channel matrices are distributed as iid unit variance
complex Gaussians (Rayleigh fading).  Furthermore, each of the receivers
is assumed to have perfect and instantaneous knowledge of
its own channel vector, i.e. ${\mathbf h}_i$.  Notice it is not
required for mobiles to know the channel of other mobiles.
Partial CSI is acquired at the transmitter via a finite rate 
feedback channel from each of the mobiles, as described below.

In the finite rate feedback model shown in Fig. \ref{fig-model} 
each receiver quantizes
its channel (with ${\bf h}_i$ assumed to be known
perfectly at the $i$-th receiver) to $B$ bits and feeds back
the bits perfectly and instantaneously to the access point,
which is assumed to have no other knowledge of the
instantaneous state of the channel.
The quantization is performed
using a vector quantization codebook that is known at the transmitter
and the receivers. A quantization codebook ${\mathcal C}$ consists of
$2^{B}$ $M$-dimensional unit norm vectors
${\mathcal C} \triangleq \{ \mathbf{w}_1, \ldots, \mathbf{w}_{2^{B}} \}$,
where $B$ is the number of feedback bits per mobile.
Similar to point-to-point MIMO systems, each receiver quantizes
its channel to the quantization vector that is closest to its
channel vector, where
closeness is measured in terms of the angle between two vectors
or equivalently the inner product
\cite{Love_Heath} \cite{Mukkavilli_MIMO}.
Thus, user $i$ computes quantization index $F_i$ according to:
\begin{eqnarray} \label{eq-quant}
F_i = \textrm {arg} \max_{j=1,\ldots,2^{B}} | {\mathbf h}_i^{\dagger} {\mathbf w}_j | = \textrm {arg} \min_{j=1,\ldots,2^{B}}
\sin^2 \left( \angle ({\bf h}_i, {\mathbf w}_j) \right).
\end{eqnarray}
and feeds this index back to the transmitter.
Note that only the direction of the channel vector is quantized, and
no information regarding the channel magnitude is conveyed to the
transmitter.  Magnitude information can be used to perform
power and rate loading across multiple channels, but this generally of
secondary concern when the number of mobiles is the same as the
number of antennas. 
If there are more users than antennas (i.e. $K > M$),
however, channel magnitude information can be used to assist with the
user selection process \cite{Yoo_Jindal_Goldsmith}.

Clearly, the choice of vector quantization codebook significantly
affects the quality of the CSI provided to the access point.
In this work, we analyze performance using random vector quantization (RVQ),
in which an ensemble of random quantization codebooks is considered. Details
of RVQ are discussed in Section \ref{sec-rvq}.

Notation: We use boldface to denote vectors and matrices
and ${\mathbf A}^{\dagger}$ refers to the conjugate transpose,
or Hermitian, of ${\mathbf A}$.  The notation
$|| {\mathbf x} ||$ refers the Euclidean norm of the
vector ${\mathbf x}$, and $\angle({\mathbf x}, {\mathbf y})$
refers to the angle between vectors ${\mathbf x}$ and ${\mathbf y}$
with the standard convention $|\cos(\angle({\mathbf x}, {\mathbf y}))|
= | {\bf x}^{\dagger} {\bf y}| / (||{\bf x}|| \cdot ||{\bf y}||)$.

\section{Background} \label{sec-background}

\subsection{Capacity Results for MIMO Broadcast Channels}
\label{sec-capacity}

In this section we summarize relevant
capacity results for the multiple-antenna broadcast channel.
When perfect CSI is available at
transmitter and receivers, the capacity region of the channel
is achieved by dirty-paper coding 
\cite{Weingarten_Steinberg_Shamai}\cite{Caire_Shamai}\cite{mimo_bc_journal}\cite{Viswanath_Tse_Journal}\cite{Yu_Cioffi}, which is a technique that 
can be used to
pre-cancel multi-user interference at the transmitter \cite{Costa}.
In this paper we study the total system
throughput, or the sum rate, which
we denote as $C_{sum}({\mathbf H}, P)$.  At high SNR,
the sum rate capacity of the MIMO BC can be approximated
as \cite{Jindal_MIMO_BC_SNR}:
\begin{eqnarray} \label{eq-dpc_approx}
C^{sum}_{TX/RX-CSI}({\mathbf H}, P) \approx
M \log \left( P \right) + c
\end{eqnarray}
where $c$ is a constant depending on the channel realization ${\mathbf H}$.
The key feature to notice is that capacity grows
\textit{linearly} as a function of $M$.  Though the $K$ (total)
receive antennas are distributed amongst $K$ receivers,
the linear growth is the same as in a $M$-transmit, $K$-receive
antenna point-to-point MIMO system, i.e. both systems
have the same \textit{multiplexing gain}.

If each of the mobiles suffers from fading according to the
same distribution and the transmitter has no instantaneous CSI,
the situation is very different.  In this scenario, the
channels of all receivers are statistically identical, and thus the
channel is degraded, in any order.  Therefore,
any codeword receiver 1 can decode can also be decoded by any other
receiver, which   implies that a TDMA strategy is optimal
\cite[Section VI]{Cover_Broadcast} \cite{Jafar_Goldsmith_Isotropic}.
Thus, the sum capacity of this channel is equal to
the capacity of the point-to-point channel from the transmitter
to any individual receiver:
\begin{equation}
C^{sum}_{RX-CSI} = E_{{\mathbf h}_1}
\left[ \log \left( 1 + \frac{P}{M} || {\mathbf h}_1 ||^2 \right) \right],
\end{equation}
and therefore the multiplexing gain of this channel is only one.
In fact, the downlink channel achieves a multiplexing gain of only
one for any fading distribution (i.e., for any distribution on the magnitude
of the channel) in which the spatial direction of each channel is 
isotropically distributed \cite{Jafar_Goldsmith_Isotropic}.
This includes a downlink channel in which users have
unequal average SNR but each suffers from spatially uncorrelated 
Rayleigh fading.

There clearly is a huge gap between the capacity of the
MIMO downlink channel with transmitter CSI (multiplexing gain of $M$)
and without transmitter CSI (multiplexing gain of $1$).
Thus, it is of interest to investigate the more practical
assumption of partial CSI at the AP.  If each of the mobiles
has perfect CSI and the AP has imperfect CSI of fixed quality,
(e.g., Rician fading with a fixed variance that is independent
of the SNR), it has recently been shown that the multiplexing gain of
the sum capacity is strictly smaller than $M$ 
\cite{Lapidoth_Shamai}\footnote{It is conjectured that the multiplexing
gain in this scenario is in fact equal to one, although this has yet
to be shown.}.  Somewhat complementary to this result, our work shows 
that the full multiplexing gain of $M$ 
can be achieved if the feedback rate (i.e., the quality of the CSI) 
is appropriately increased as a function of SNR.

\subsection{Downlink Precoding}
Though dirty paper coding is capacity achieving for the MIMO
broadcast channel, the technique requires considerable complexity and
practical implementations are still being actively pursued
\cite{Zamir2002,Erez2004,Philosof2003}.  As a result,
simpler downlink transmission schemes are of obvious interest.
One such scheme is downlink beamforming\footnote{Downlink
beamforming is also referred to as linear precoding or
space-division multiple access (SDMA).}, which
incurs a rate/power loss relative to DPC but
achieves the same multiplexing gain of $M$.  In order to
implement this scheme, the transmitter multiplies the
symbol intended for each receiver by a beamforming
vector and transmits the sum of these vector signals.
Let $s_i$ denote the scalar symbol intended for the $i$-th receiver,
and let ${\bf v}_i$ denote the corresponding unit norm beamforming vector.
The transmitted signal is then given by:
\begin{eqnarray}
{\mathbf x} =  \sum_{j=1}^K {\mathbf v}_j s_j,
\end{eqnarray}
The received signal at user $i$ is therefore given by:
\begin{eqnarray}
y_i &=& {\mathbf h}_i^{\dagger} {\mathbf x} + n_i
= \sum_{j=1}^K {\mathbf h}_i^{\dagger} {\mathbf v}_j s_j + n_i,
\end{eqnarray}
and the SINR at mobile $i$ is:
\begin{eqnarray} \label{eq-downlink_sinr}
SINR_i = \frac { \frac{P}{M} | {\mathbf h}_i^{\dagger} {\mathbf v}_i |^2 }
{1 + \sum_{j \ne i}  \frac{P}{M} | {\mathbf h}_i^{\dagger} {\mathbf v}_j |^2 },
\end{eqnarray}
under the assumption that each of the symbols has power $P/M$,
and that no interference cancellation is performed at the mobiles.
Note that the capacity-achieving strategy is similar to
downlink beamforming with the addition of a pre-coding step at the
transmitter which leads to the elimination of some of
the multi-user interference terms in the SINR expression.

The performance of downlink beamforming clearly depends
on the choice of beamforming vectors, but the problem
of determining the sum rate maximizing beamforming vectors is
generally very difficult. One simple choice
of beamforming vectors are the zero-forcing vectors,
which are chosen such that
no multi-user interference is experienced at any of
the receivers.  This can be done by choosing
the beamforming vector of user $i$ orthogonal to the channel
vectors of all other users, i.e., by choosing ${\bf v}_i$
orthogonal to ${\bf h}_j$ for all $i \ne j$.  It is easily seen
that the zero-forcing beamforming vectors are simply the normalized
columns of the inverse of the concatenated channel matrix ${\bf H}$.
If such beamforming vectors are used, the received
signal at the $i$-th mobile reduces to:
\begin{eqnarray}
y_i &=& {\mathbf h}_i^{\dagger} {\mathbf x} + n_i
= \sum_{j=1}^K {\mathbf h}_i^{\dagger} {\mathbf v}_j s_j + n_i =
{\mathbf h}_i^{\dagger} {\mathbf v}_i s_i + n_i
\end{eqnarray}
because ${\bf h}_i^{\dagger} {\bf v}_j = 0$ for all $j \ne i$ by
construction.  Since all interference has been eliminated,
the corresponding SNR is given as
$SNR_i = \frac{P}{M} | {\mathbf h}_i^{\dagger} {\mathbf v}_i |^2$.
In fact, zero-forcing is optimal amongst all 
downlink beamforming strategies at
asymptotically optimal at high SNR \cite{Jindal_MIMO_BC_SNR}.

Since zero-forcing creates
$M$ independent and parallel channels, the resulting
multiplexing gain is equal to $M$, which is the same
as for the capacity-achieving DPC strategy.  Zero forcing
does however, incur a rate loss (or alternatively, a power loss)
relative to capacity.  At high SNR, the power loss of
zero forcing relative to DPC converges to
$\frac{3 \log_2 e}{M} \sum_{j=1}^{M-1} \frac{j}{M-j}$ dB,
which is approximately equal to $3 \log_2 M$ dB
\cite{Jindal_MIMO_BC_SNR}.
Clearly, the transmitter must have perfect channel knowledge
in order to choose the zero-forcing beamforming vectors.
If there is any imperfection in this knowledge, there inevitably
will be some multi-user interference, which leads to 
performance degradation.

\subsection{Random Vector Quantization} \label{sec-rvq}
In this work we use \textit{random vector quantization} (RVQ),
in which each of the $2^{B}$ quantization
vectors is independently chosen from the isotropic distribution
on the $M$-dimensional unit sphere.  
We analyze performance
averaged over all such choices of random codebooks, in addition
to averaging over the fading distribution.  Random codebooks are used
because the optimal vector quantizer for this problem is
not known in general, and known bounds are rather loose.
RVQ, on the other hand, is very amenable to analysis and also 
performs measurably close to optimal quantization, as is
shown in Section \ref{sec-rvq_opt}.  Note that each
receiver is assumed to use a different and independently
generated quantization codebook; if a common codebook was used,
there would be a non-zero probability that multiple users
return the same quantization vector, which complicates
transmission.

Random vector quantization was first used to analyze the performance of
CDMA and point-to-point MIMO channels with finite rate feedback, and
has been shown to be asymptotically optimal in the large system
limit (e.g., infinitely many antennas) 
\cite{Santipach_Honig_CDMA}\cite{Santipach_Honig}.
There has also been very recent work characterizing
the error performance of point-to-point MISO (multiple-input,
single-output) systems utilizing RVQ \cite{Yeung_Love}.

We now review some basic results on RVQ from 
\cite{Yeung_Love} that will
be useful in later derivations.  As stated earlier, the
quantization vectors are iid isotropic vectors on the
$M$-dimensional unit sphere, as are the channel
directions $\tilde{\bf h}_i \triangleq
\frac{ {\bf h}_i } {|| {\bf h}_i ||}$
due to the assumption of iid Rayleigh fading.
The most important quantity of interest is the statistical distribution
of the quantization error.  In order to determine this,
first consider the inner product between a channel vector
and a quantization vector:
\begin{eqnarray*}
|{\mathbf w}_j^{\dagger} {\mathbf h}_i|^2  = ||{\bf h}_i||^2
\left| {\mathbf w}_j^{\dagger} \tilde{\bf h}_i \right|^2
= ||{\bf h}_i||^2  \cos^2 \left( \angle (\tilde{\bf h}_i, {\mathbf w}_j) \right).
\end{eqnarray*}
Because $\tilde{\bf h}_i$ and ${\bf w}_j$ are independent
isotropic vectors, the quantity
$| {\mathbf w}_j^{\dagger} \tilde{\bf h}_i |^2 =
\cos^2 \left( \angle (\tilde{\bf h}_i, {\mathbf w}_j) \right)$
is beta distributed with parameters $1$ and $M-1$, 
and $\sin^2 \left( \angle (\tilde{\bf h}_i, {\mathbf w}_j) \right) = 1 - 
\cos^2 \left( \angle ({\bf h}_i, {\mathbf w}_j) \right)$
is beta distributed with parameters $M-1$ and $1$. Thus the CDF of 
$X= \sin^2 \left( \angle (\tilde{\bf h}_i, {\mathbf w}_j) \right)$
is given by $\textrm{Pr}(X \leq x) = x^{M-1}$.

Let $\hat{\bf h}_i$
denote the quantization of the vector ${\bf h}_i$, i.e. the
solution to (\ref{eq-quant}).  Since the quantization
vectors are independent, the quantization error
$Z \triangleq \sin^2 \left( \angle (\tilde{\bf h}_i, \hat{\bf h}_i) \right)$
is the minimum of $2^{B}$ independent beta $(M-1,1)$ random variables,
and the CCDF of $Z$ is given by $\textrm{Pr}(Z \geq z) = (1-z^{M-1})^{2^{B}}$
\cite[Lemma 1]{Yeung_Love}.
The expectation of this quantity has been computed in 
closed form \cite{Yeung_Love}:
\begin{eqnarray} \label{eq-expected_quant}
E_{{\bf H}, {\mathcal W}} \left[
\sin^2 \left( \angle (\tilde{\bf h}_i, \hat{\bf h}_i) \right) \right]
= 2^{B} \cdot \beta \left(2^{B}, \frac{M}{M-1}\right).
\end{eqnarray}
Here we use $\beta(\cdot)$ to denote the beta function, which is defined
in terms of the gamma function as
$\beta(x,y) = \frac{ \Gamma(x) \Gamma(y)}{ \Gamma(x+y)}$ \cite{Davis_Gamma}.
The gamma function is the extension of the factorial function
to non-integers, and satisfies the fundamental properties
$\gamma(n) = (n-1)!$ for positive integers and 
$\Gamma(x+1)=x \Gamma(x)$ for all $x$ \cite{Davis_Gamma}.
While the derivation of the expectation in 
\cite{Yeung_Love} depends on the Pochmann
symbol, this result can alternatively be derived using an
integral representation of the beta function, as shown in
Appendix \ref{app-quant_error}.
Furthermore, a simple extension of inequalities given in \cite{Yeung_Love}
gives a strict upper bound to the expected quantization error:
\begin{lemma} \label{lem-quant_upper}
The expected quantization error can be upper bounded as:
\begin{eqnarray*} 
E_{{\bf H}, {\mathcal W}} \left[
\sin^2 \left( \angle (\tilde{\bf h}_i, \hat{\bf h}_i) \right) \right]
< 2^{- \frac{B}{M-1}}.
\end{eqnarray*}
\end{lemma}
\begin{proof} See Appendix \ref{app-quant_upper}.
\end{proof}

\subsection{Point-to-Point MISO Systems with Finite Rate Feedback}
\label{sec-miso}
In this section we briefly review some basic results
on point-to-point MISO systems (i.e., $M$ transmit antennas,
single receive antenna, which is equivalent to the given
system model with $K=1$) with finite rate feedback, under the
assumption of iid Rayleigh fading.
If the transmitter has perfect CSI, it is well known that the
optimum transmission strategy is to beamform along the
channel vector ${\bf h}$ \cite{Telatar} and the corresponding (ergodic)
capacity is
$C_{\textrm{CSIT}}(P) =
E_{{\bf h}} \left[ \log \left(1 + P ||{\bf h}||^2 \right) \right]$.
If the transmitter has no CSIT and only has knowledge of
the fading distribution, the optimum transmission strategy is
to transmit independent and equal power signals from each of the
$M$ transmit antennas, and the corresponding capacity is
$C_{\textrm{no-CSIT}}(P) =
E_{{\bf h}} \left[ \log \left(1 + \frac{P}{M} ||{\bf h}||^2 \right) \right]$.
Clearly $C_{\textrm{no-CSIT}}(P) = C_{\textrm{CSIT}}(\frac{P}{M})$,
which corresponds to a $10 \log_{10} M$ dB shift in the capacity curve.
Thus the lack of CSIT leads to a $10 \log_{10} M$ dB
SNR loss relative to perfect CSIT.

\begin{figure}
\begin{center}
\centerline{\epsfig{file=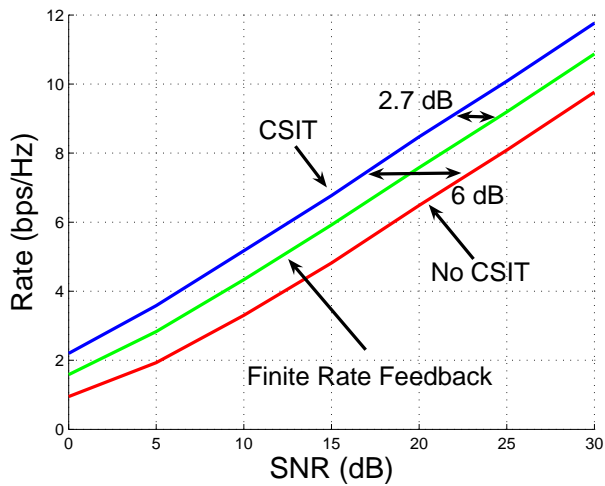, width=90mm}}
\end{center}
\caption{4 x 1 MISO System with CSIT, No CSIT, \& Feedback}
\label{fig-miso}
\end{figure}

Providing the transmitter with partial CSIT via a finite
rate feedback channel can be used to reduce this SNR loss.
If the transmitter acquires CSIT through the
finite rate feedback channel, an optimal or nearly optimal strategy is to
beamform in the direction of the quantization vector\footnote{Conditions
for the optimality of beamforming along the
quantization direction are provided in \cite{Jafar_Srinivasa}.
Though these conditions are difficult to analytically compute for
most quantization codebooks, it is generally well accepted that
beamforming performs extremely close to capacity.}.  The
average rate achieved with this strategy assuming RVQ is used is given by:
\begin{eqnarray*}
R_{FB}(P) &=& E_{{\bf h}, {\mathcal W}}
\left[ \log \left(1 + P ||{\bf h}||^2
\cos^2 \left( \angle ({\bf h}, \hat{\bf h}) \right) \right) \right] \\
&\approx& E_{{\bf h}}
\left[ \log \left(1 + P ||{\bf h}||^2 (1 - 2^{- \frac{B}{M-1}})
\right) \right],
\end{eqnarray*}
where we have based the approximation on the upper bound given in 
Lemma \ref{lem-quant_upper}, which is numerically very accurate.
Thus, the use of limited feedback leads to an SNR degradation
of approximately $10 \log_{10} (1 - 2^{- \frac{B}{M-1}})$ dB
relative to perfect CSIT.
Note that this approximation agrees with the expression
derived for an asymptotically large number of transmit antennas
in \cite{Santipach_Honig}.  If $B=M-1$ for example,
a finite rate feedback system is expected to perform within about
3 dB of a perfect CSIT system.  The capacity with
CSIT, no CSIT, and finite rate feedback with $B=M-1=3$ is shown
for a $4 \times 1$ MISO system in Fig. \ref{fig-miso}.
Notice that there is a 6 dB gap between the CSIT and no CSIT
curves, while the finite rate feedback curve is 2.7 dB
from the perfect CSIT curve, which is quite close to 
our approximation of 3 dB.  

The key point to take from this discussion is that the
feedback load need not be increased as a function of SNR in order
to maintain a constant power or rate gap relative to the perfect
CSIT capacity curve.  This is perhaps not surprising,
since the amount of feedback only affects the
$\cos^2 \left( \angle ({\bf h}, \hat{\bf h}) \right)$ term
and thus leads to only a SNR degradation.
Furthermore, also note that the multiplexing gain (i.e., the slope of the
capacity curve) is not affected by the level of CSIT.
For the MIMO downlink channel, however, the multiplexing gain
of a $M$-transmit antenna, $K$ user system is $\min(N,K)$
when there is perfect CSIT, but is only $1$ if there is no
CSIT.  Clearly such a channel will be considerably more
sensitive to the accuracy of the CSIT obtained through the
finite rate feedback channels.  

\section{Downlink Precoding with Finite Rate Feedback}
\label{sec-sys_description}
After the transmitter has received feedback bits from each of
the $K$ receivers, an appropriate multi-user transmission strategy
must be chosen.  We propose using \textit{zero-forcing beamforming}
based on the channel quantizations available at the transmitter.
Zero-forcing (ZF) is a low complexity transmission scheme that can be
implemented by a simple linear precoder, and its performance
is optimal amongst the set of all linear precoders at asymptotically
high SNR \cite{Jindal_MIMO_BC_SNR}.  
We later provide numerical results describing 
 the performance of \textit{regularized}
zero-forcing precoding \cite{Peel2005}, which outperforms
pure zero-forcing at low SNR's but is equivalent to ZF at
asymptotically high SNR.

Note that dirty paper coding cannot be directly applied
in this scenario because of the imperfection in CSIT.
In order to implement DPC, the transmitter in fact requires
knowledge of the multi-user interference at the receiver, and
not at the transmitter.  The received interference clearly
depends on the channel state, which is not known
perfectly at the transmitter in the finite rate feedback model.
The transmitter could estimate the received interference based
on the available channel quantization, but even this
estimated interference cannot be cancelled perfectly using DPC
due to the transmitter's imperfect knowledge of the received
signal power.  In order to implement DPC, the transmitter must
also know the received SNR (in the absence of any interference)
in order to properly select the \textit{inflation factor}, which
is a key component in the dirty paper coding implementation
\cite{Zamir2002}.
Since the SNR also depends on the channel realization,
the transmitter only has an imperfect estimate of the
SNR and thus cannot properly select the inflation factor.

When the transmitter has perfect CSI, zero-forcing
can be used to completely eliminate multi-user interference
by precoding transmission by the inverse of the channel matrix
${\bf H}$. This creates a parallel, non-interfering channel to
each of the $M$ receivers, and thus leads to a multiplexing gain of $M$.
In the finite rate feedback setting, the imperfection in CSIT
makes it impossible to completely eliminate all multi-user interference,
but a zero-forcing based strategy can still be quite effective.
Since the transmitter only has knowledge of the channel quantizations
but does not have any information regarding the magnitude or
spatial direction of the quantization error, a reasonable approach
to take is to select beamforming vectors according to the
zero-forcing criterion based on the channel quantizations.

Let $\hat{\bf h}_i$ refer to the quantized version of the mobile $i$'s
channel. These quantized vectors are compiled into a
matrix: $\hat{\bf H} = \left[ \begin{array}{cccc}
\hat{\bf h}_1 & \hat{\bf h}_2 & \cdots & \hat{\bf h}_M \end{array} \right]^{\dagger}$.
The matrix ${\hat{\bf H}}$ is the estimate of the channels, upon which
zero-forcing is performed.  Thus, the beamforming vectors
are chosen to be the normalized columns of the matrix ${\hat{\bf H}}^{-1}$.
If equal power $\frac{P}{M}$ is used for each of the data streams,
the received SINR at the $i$-th mobile is given by (\ref{eq-downlink_sinr}):
\begin{eqnarray} \label{eq-sinr}
SINR_i = \frac { \frac{P}{M} | {\mathbf h}_i^{\dagger} {\mathbf v}_i |^2 }
{1 + \sum_{j \ne i}  \frac{P}{M} | {\mathbf h}_i^{\dagger} {\mathbf v}_j |^2 }.
\end{eqnarray}
Since the beamforming vectors are chosen orthogonal to the
channel quantizations and not the actual channel realizations,
the interference terms in the denominator of the SINR expression
are not zero.  However, these terms directly depend
on the quantization error and thus can be analyzed using the
statistics of random vector quantization.

We study long-term average throughput (over both the fading
distribution and RVQ), and thus
the rate of transmission to User $i$ is equal to
$R_i =  E_{\mathbf H, {\mathcal W}} \left[ \log_2( 1 + SINR_i) \right]$
if Gaussian inputs are used.  
By symmetry, the system throughput is given by:
\begin{eqnarray*}
R_{FB}(P) \triangleq M 
E_{\mathbf H, {\mathcal W}} \left[ \log_2( 1 + SINR) \right].
\end{eqnarray*}
For a system that achieves a throughput of $R(P)$ (where $P$
is the SNR, or power constraint), the multiplexing gain
is defined as:
\begin{eqnarray} \label{eq-multiplexing}
r = \lim_{P \rightarrow \infty} \frac {R(P)}{ \log_2 (P)}.
\end{eqnarray}

\section{Throughput Analysis} \label{sec-throughput}
In this section we analyze the throughput of a feedback-based
zero-forcing system.  We first state some useful preliminary calculations,
and then study the achieved throughput for fixed and
increasing feedback levels.

\subsection{Preliminary Calculations}
In this section we prove a few useful results regarding the distribution
of terms in the SINR expression in (\ref{eq-sinr}).
For the remainder of this paper, we use $\tilde{\bh}_i$ to denote
the normalized channel vector, i.e., $\tilde{\bh}_i = 
{\bh}_i / ||{\bh}_i||$.  Using this notation, we can rewrite
the SINR as:
\begin{eqnarray} 
SINR_i = \frac { \frac{P}{M} ||{\bh}_i||^2 
| \tilde{\mathbf h}_i^{\dagger} {\mathbf v}_i |^2 }
{1 + \sum_{j \ne i}  \frac{P}{M} ||{\bh}_i||^2  
| \tilde{\mathbf h}_i^{\dagger} {\mathbf v}_j |^2 }.
\end{eqnarray}

We first characterize the numerator of this expression, i.e., the
received signal power:
\begin{lemma} \label{lem-bf_vector}
The beamforming vector ${\bf v}_i$ is isotropically distributed
in $\mathbb{C}^M$ and is independent of the channel direction
 $\tilde{\bf h}_i$ as well as the channel quantization $\hat{\bf h}_i$.
\end{lemma}
\begin{proof}
By the zero-forcing procedure, ${\bf v}_i$ is chosen
in the nullspace of $\{ \hat{\bf h}_j \}_{j \ne i}$.  Since
RVQ is used and the channel directions $\{ \tilde{\bf h}_j \}_{j \ne i}$
are independent isotropic vectors, the channel quantizations
$\hat{\bh}_1, \ldots, \hat{\bh}_M$ are mutually independent
isotropically distributed vectors.  Thus the nullspace of
$\{ \hat{\bf h}_j \}_{j \ne i}$ is an isotropically distributed
direction in $\mathbb{C}^M$, independent of either $\tilde{\bf h}_i$ or
$\hat{\bf h}_i$.
\end{proof}
Clearly the same argument holds if the transmitter performs 
zero-forcing on the basis of perfect CSIT 
(i.e., $\hat{\bf h}_i = \tilde{\bf h}_i$).

Next we characterize the interference terms that appear in the
denominator of the SINR expression:
\begin{lemma} \label{lem-int_terms}
The random variable 
$| \tilde{\mathbf h}_i^{\dagger} {\mathbf v}_j |^2$ for any $i \ne j$ 
is equal to the product of the quantization error 
$\sin^2 \left( \angle (\tilde{\bf h}_i, \hat{\bf h}_i) \right)$
and an independent beta $(1,M-2)$ random variable.
\end{lemma}
\begin{proof}
Without loss of generality, consider $i=2$ and $j=1$, i.e.
the term $| \tilde{\mathbf h}_2^{\dagger} {\mathbf v}_1 |^2$.  The vector
${\mathbf v}_1$ is chosen in the nullspace of 
$\hat{\bh}_2, \ldots, \hat{\bh}_M$, each of which is an independent
isotropically distributed vector.  Therefore, ${\mathbf v}_1$ is
isotropically distributed within the $(M-1)$-dimensional 
nullspace of $\hat{\bh}_2$.  Now consider the normalized
channel vector $\tilde{\bh}_2$.  Since RVQ is used, the quantization
error has no preferential direction, i.e. the error
is isotropically distributed in $\mathbb{C}^M$.  Thus, conditioned
on the magnitude of the quantization error
$a \triangleq \sin^2 \left( \angle (\tilde{\bf h}_i, \hat{\bf h}_i) \right)$,
the channel direction can be written as the sum of two vectors, one
in the direction of the quantization, and the other isotropically
distributed in the nullspace of the quantization:
$\tilde{\bh}_2 = \sqrt{1 - a} \hat{\bh}_2 + \sqrt{a} {\mathbf s}$,
where ${\mathbf s}$ is isotropically distributed in the
nullspace of $\hat{\bh}_2$, and is independent of $a$.  
Therefore, the random variable $\tilde{\bh}_2$ can be written as:
\begin{eqnarray*}
\tilde{\bh}_2 = (\sqrt{1 - Z}) \hat{\bh}_2 + \sqrt{Z} {\mathbf s},
\end{eqnarray*}
where ${\bf s}$ and $Z$ are independent, with
${\bf s}$ isotropically distributed in the nullspace of $\hat{\bh}_2$
and $Z$ distributed according to the quantization error distribution,
i.e., the minimum of $2^{B}$ beta $(M-1,1)$ random variables,
as described in Section \ref{sec-rvq}.

The inner product of $\tilde{\bh}_2$ and ${\mathbf v}_1$ is then given by:
\begin{eqnarray*}
| \tilde{\mathbf h}_2^{\dagger} {\mathbf v}_1 |^2 &=&
(1 - Z) | \hat{\bh}_2^{\dagger} {\mathbf v}_1|^2  + Z |{\mathbf s}^{\dagger} 
{\mathbf v}_1|^2 \\
&=& Z |{\mathbf s}^{\dagger} {\mathbf v}_1|^2.
\end{eqnarray*}
Since ${\bf s}$ and ${\bf v}_1$ are iid isotropic vectors in
the $(M-1)$-dimensional nulllspace of $\hat{\bh}_2$, the 
quantity $|{\mathbf s}^{\dagger} {\mathbf v}_1|^2$ is beta $(1,M-2)$
distributed, and is independent of $Z$.
\end{proof}
Since a beta random variable has support $[0,1]$, we have
\begin{eqnarray} \label{eq-int_upper_bound}
| \tilde{\mathbf h}_i^{\dagger} {\mathbf v}_j |^2 \leq 
\sin^2 \left( \angle (\tilde{\bf h}_i, \hat{\bf h}_i) \right),~~~~
\forall i \ne j,
\end{eqnarray}
i.e., the interference from any single user is no
larger than the quantization error.
When $M=2$, we clearly have 
$| \tilde{\mathbf h}_i^{\dagger} {\mathbf v}_j |^2 = 
\sin^2 \left( \angle (\tilde{\bf h}_i, \hat{\bf h}_i) \right)$, and no
beta random variable is needed.

Finally, a derivation of the expectation of the logarithm of the 
quantization error, which is useful in a few subsequent theorems, is given:
\begin{lemma} \label{lem-exp_log}
The expectation of the logarithm of the quantization error is given by:
\begin{eqnarray*}
E_{\mathbf H, {\mathcal W}} \left[ \log_2 \left(
\sin^2 ( \angle(\tilde{\mathbf h}_i, \hat{{\bf h}}_i) ) \right) \right] = 
- \frac{\log_2 e}{M-1} \sum_{k=1}^{2^{B}} \frac{1}{k}.
\end{eqnarray*}
Furthermore, this quantity can be bounded as:
\begin{eqnarray*}
 \frac{B}{M-1} \leq - E_{\mathbf H, {\mathcal W}} \left[ \log_2 \left(
\sin^2 ( \angle(\tilde{\mathbf h}_i, \hat{{\bf h}}_i) ) \right) \right] \leq 
\frac{B+\log_2 e}{M-1}.
\end{eqnarray*}
\end{lemma}

\begin{proof}
See Appendix \ref{app-log}.
\end{proof}

\subsection{Fixed Feedback Quality}

We now analyze the average throughput achieved by the
proposed zero-forcing scheme, and quantify the performance
degradation 
as a function of the feedback rate.
In order to study the performance loss, we define the
rate gap $\Delta R (P)$ 
to be the difference between the per mobile throughput achieved by
perfect CSIT-ZF and finite-rate feedback based ZF:
\begin{eqnarray*}
\Delta R (P) &\triangleq& \frac{1}{M} [R_{ZF}(P) - R_{FB}(P)].
\end{eqnarray*}
In this expression $R_{ZF}(P)$ refers to the throughput achieved by 
perfect CSIT-based zero-forcing (i.e., $\hat{\bf h}_i = \tilde{\bf h}_i$), 
which is given by:
\begin{eqnarray*}
R_{ZF}(P) =  M E_{{\bf H}} \log_2 \left( 1 + \frac{P}{M} 
| {\mathbf h}_i^{\dagger} {\mathbf v}_{ZF,i} |^2 \right),
\end{eqnarray*}
where each beamforming vector ${\mathbf v}_{ZF,i}$ is chosen orthogonal
to $\{ {\bf h}_j \}_{j \ne i}$.

\begin{theorem} \label{thm-power-loss}
Finite rate feedback with $B$ feedback bits per mobile
incurs a throughput loss relative to perfect CSIT zero forcing that
can be upper bounded by:
\begin{eqnarray*}
\Delta R (P) < \log_2 \left( 1 + P \cdot 2^{-\frac{B}{M-1}} \right).
\end{eqnarray*}
\end{theorem}

\begin{proof}
The rate gap can be upper bounded as:
\begin{eqnarray*}
\Delta R (P) 
&=&  E_{{\bf H}} \left[ \log_2 \left(1 + \frac{P}{M} |{\bf h}_i^{\dagger}
{\bf v}_{ZF,i}|^2 \right) \right] -
E_{{\bf H},W} \left[ \log_2 \left(1 +
\frac { \frac{P}{M} | {\mathbf h}_i^{\dagger} {\mathbf v}_i |^2 }
{1 + \sum_{j \ne i}  \frac{P}{M} | {\mathbf h}_i^{\dagger} {\mathbf v}_j |^2 }
\right) \right] \\
&=&  E_{{\bf H}} \left[ \log_2 \left(1 + \frac{P}{M} |{\bf h}_i^{\dagger}
{\bf v}_{ZF,i}|^2 \right) \right] -
E_{{\bf H},W} \left[ \log_2 \left(
1 + \frac{P}{M} | {\mathbf h}_i^{\dagger} {\mathbf v}_i |^2  + \sum_{j \ne i}  \frac{P}{M} | {\mathbf h}_i^{\dagger} {\mathbf v}_j |^2 \right) \right] \\
&& + E_{{\bf H},W} \left[ \log_2 \left(
1 + \sum_{j \ne i}  \frac{P}{M} | {\mathbf h}_i^{\dagger} {\mathbf v}_j |^2 \right) \right] \\
&\stackrel{(a)}{\leq}&
E_{{\bf H}} \left[ \log_2 \left(1 + \frac{P}{M} |{\bf h}_i^{\dagger}
{\bf v}_{ZF,i}|^2 \right) \right] -
E_{{\bf H},W} \left[ \log_2 \left(
1 + \frac{P}{M} | {\mathbf h}_i^{\dagger} {\mathbf v}_i |^2  \right) \right] 
\\ && + E_{{\bf H},W} \left[ \log_2 \left(
1 + \sum_{j \ne i}  \frac{P}{M} | {\mathbf h}_i^{\dagger} {\mathbf v}_j |^2 \right) \right] \\
&\stackrel{(b)}{=}&
E_{{\bf H},W} \left[ \log_2 \left(1 + \sum_{j \ne i}  \frac{P}{M} | {\mathbf h}_i^{\dagger} {\mathbf v}_j |^2 \right) \right],
\end{eqnarray*}
where (a) follows because 
$\sum_{j \ne i}  \frac{P}{M} | {\mathbf h}_i^{\dagger} {\mathbf v}_j |^2  \geq 0$ and 
$\log(\cdot)$ is a monotonically increasing function.  To get
(b), note that ${\bf v}_{ZF,i}$ and ${\bf v}_i$
are each isotropically distributed unit vectors, independent
of ${\bf h}_i$, by Lemma \ref{lem-bf_vector}, which implies
$E_{{\bf H}} \left[ \log_2 \left(1 + \frac{P}{M} |{\bf h}_i^{\dagger}
{\bf v}_{ZF,i}|^2 \right) \right] =
E_{{\bf H},W} \left[ \log_2 \left(
1 + \frac{P}{M} | {\mathbf h}_i^{\dagger} {\mathbf v}_i |^2  \right) \right]$.
Applying Jensen's inequality and
exploiting the independence of the channel norm (which satisfies
$E[||{\bh}_i||^2]  = M$) and channel direction, we get:
\begin{eqnarray*}
\Delta R(P) &\leq&  \log_2 \left( 1 +  \frac{P}{M} (M-1) E[||{\bh}_i||^2] 
E[ | \tilde{\bf h}_i^{\dagger} {\bf v}_j |^2] \right) \\
&=& \log_2 \left( 1 +  P (M-1) E[| \tilde{\bf h}_i^{\dagger} {\bf v}_j |^2] \right). 
\end{eqnarray*}
By Lemma \ref{lem-int_terms}, the term
$E[| \tilde{\bf h}_i^{\dagger} {\bf v}_j |^2]$ is the product of
the expectation of the quantization error and the expectation of a 
beta$(1,M-2)$ random variable, which is equal to $\frac{1}{M-1}$.
Using Lemma \ref{lem-quant_upper} we have:
\begin{eqnarray*}
\Delta R(P) &\leq& \log_2 \left( 1 +  P \cdot E[ \sin^2 ( \angle(\tilde{\mathbf h}_i, \hat{{\bf h}}_i) )] \right) \\
&<& \log_2 \left( 1 + P \cdot 2^{-\frac{B}{M-1}} \right).
\end{eqnarray*}
\end{proof}

The most important feature to notice is that the rate loss is an
increasing function of the system SNR ($P$), which can be explained by the
linear relationship between $P$ and the multi-user interference power. 
This intuition motivates the following result, which shows that
a finite-rate feedback system with fixed feedback quality is interference-limited
at high SNR:
\begin{theorem} \label{thm-fixed_bits}
The throughput achieved by finite-rate feedback-based zero-forcing
with a fixed number of feedback bits per mobile $B$ is bounded as
the SNR is taken to infinity:
\begin{eqnarray*}
R_{FB}(P) &\leq& M\left( 1 + \frac{B + \log_2 e}{M-1} + \log_2 e + \log_2(M-2) \right).
\end{eqnarray*}
\end{theorem}
\begin{proof}
Consider the following upper bounds to the throughput $R_{FB}(P)$:
\begin{eqnarray*}
\frac{1}{M} R_{FB}(P) &=& E_{\mathbf H, {\mathcal W}} \left[ \log_2 
\left( 1 + \frac { \frac{P}{M} | {\mathbf
h}_i^{\dagger} {\mathbf v}_i |^2 } {1 + \sum_{j \ne i}
\frac{P}{M} | {\mathbf h}_i^{\dagger} {\mathbf v}_j |^2 } \right) \right] \\
 &\leq& E_{\mathbf H, {\mathcal W}} \left[ \log_2  \left( 1 + 
\frac { ||{\mathbf h}_i||^2 | \tilde{\mathbf h}_i^{\dagger} {\mathbf v}_i |^2 } {\sum_{j \ne i} ||{\mathbf h}_i||^2
 | \tilde{\mathbf h}_i^{\dagger} {\mathbf v}_j |^2 } \right) \right] \\
&\stackrel{(a)}{\leq}& E_{\mathbf H, {\mathcal W}} \left[ \log_2 
\left( 1 + \frac { | \tilde{\mathbf h}_i^{\dagger} {\mathbf v}_i |^2 } 
{ | \tilde{\mathbf h}_i^{\dagger} {\mathbf v}_j |^2 } \right) \right] \\
&=& E_{\mathbf H, {\mathcal W}} \left[ \log_2 \left(
| \tilde{\mathbf h}_i^{\dagger} {\mathbf v}_j |^2 + 
| \tilde{\mathbf h}_i^{\dagger} {\mathbf v}_i |^2 \right) \right]-
E_{\mathbf H, {\mathcal W}} \left[ \log_2 \left(
| \tilde{\mathbf h}_i^{\dagger} {\mathbf v}_j |^2  \right) \right] \\
&\stackrel{(b)}{\leq}& 1 -
E_{\mathbf H, {\mathcal W}} \left[ \log_2 \left(
| \tilde{\mathbf h}_i^{\dagger} {\mathbf v}_j |^2  \right) \right].
\end{eqnarray*}
where in (a) we consider only one of the multi-user interference 
terms, and (b) uses the fact that 
$| \tilde{\mathbf h}_i^{\dagger} {\mathbf v}_j | \leq 1$ and
$| \tilde{\mathbf h}_i^{\dagger} {\mathbf v}_i | \leq 1$. 
Since  $|\tilde{\mathbf h}_i^{\dagger} {\mathbf v}_j |^2 $ is the product
of the quantization error and a beta random variable (Lemma \ref{lem-int_terms}), 
we have
\begin{eqnarray*}
- E_{\mathbf H, {\mathcal W}} \left[ \log_2 \left(
| \tilde{\mathbf h}_i^{\dagger} {\mathbf v}_j |^2  \right) \right]
&=& - E_{\mathbf H, {\mathcal W}} \left[ \log_2 \left(
\sin^2 ( \angle(\tilde{\mathbf h}_i, \hat{{\bf h}}_i) ) \right) \right] 
- E_Y \left[ \log_2 Y \right] \\
&=& \frac{\log_2 e}{M-1} \sum_{k=1}^{2^{B}} \frac{1}{k} + 
\log_2 e \sum_{k=1}^{M-2} \frac{1}{k} \\
&\leq& \frac{B + \log_2 e}{M-1}  + \log_2(M-2) + \log_2 e,
\end{eqnarray*}
where we have used Lemma \ref{lem-exp_log} as well as the easily
verifiable fact that
$E[-\log_2 Y] = (\log_2 e) \sum_{k=1}^{M-2} \frac{1}{k} \leq
\log_2(M-2) + \log_2 e$ when  
$Y$ is beta $(1,M-2)$.  Plugging this expression into the
upper bound on $R_{FB}(P)$ gives the result.
\end{proof}

\begin{figure}
\begin{center}
\centerline{\epsfig{file=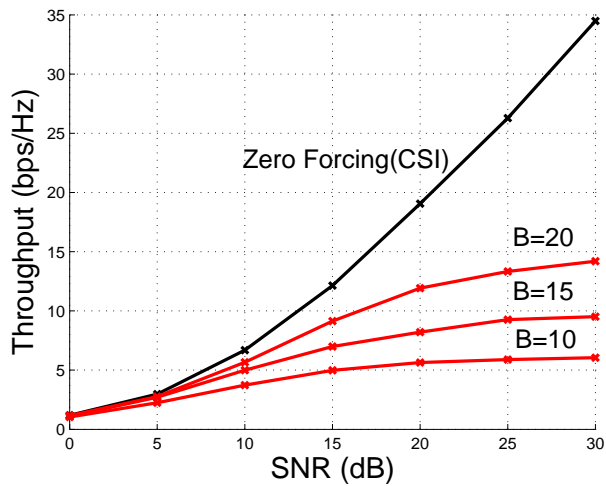, width=90mm}}
\end{center}
\caption{5 $\times$ 5 Channel with Fixed \# Feedback Bits}
\label{fig1}
\end{figure}

Regardless of how many feedback bits ($B$) are used, the system 
eventually becomes interference limited
because interference and signal power both scale linearly
with $P$.
In Fig. \ref{fig1}
the performance of a 5 antenna, 5 user system with 10, 15, and
20 feedback bits per mobile is shown.  When the SNR is
small, limited feedback performs nearly as well as
zero-forcing.  However, as the SNR is increased, the
limited feedback system becomes interference limited
and the rates converge to an upper limit, as expected.
Although the upper bound in Theorem \ref{thm-fixed_bits}
is quite loose in general, it does correctly predict the
roughly linear dependence of the limiting throughput and $B$.

Notice that this interference-limited behavior can easily be 
avoided by reverting to a TDMA strategy, but this only provides 
a multiplexing gain of one.  However, as validated by numerical results
in Section \ref{sec-comparison}, TDMA is preferable
to feedback-based zero forcing at high SNR's if $B$ is kept fixed.


\subsection{Increasing Feedback Quality}

In the previous section we studied systems with fixed feedback
rates, and showed that the throughputs of such systems are bounded.
In this section we show that this interference-limited behavior
can be avoided by scaling the feedback rate linearly with the SNR (in dB).
In fact, if the feedback rate is scaled at the appropriate rate, the
full multiplexing gain of $M$ is achievable.  In addition to
achieving the full multiplexing gain, it is also desirable to
maintain a constant rate offset between 
the rates achievable with zero-forcing with perfect CSI and
with finite-rate feedback.  Note that if a bounded rate gap
is maintained, the full multiplexing gain is also achieved.
The following
theorem specifies a sufficient scaling of feedback bits to maintain
a bounded rate gap:

\begin{theorem} \label{thm-scaling}
In order to maintain a rate offset no larger than $\log_2 b$
(per user) between zero forcing with perfect CSI and 
with finite-rate feedback, it is sufficient to scale the
number of feedback bits per mobile according to:
\begin{eqnarray}  \label{eq-N_FB_thm}
B &=& (M-1) \log_2 P -  (M-1)\log_2(b-1) \\
&\approx& \frac{M-1}{3} P_{dB} -  (M-1)\log_2(b-1). \label{eq-scaling_thm}
\end{eqnarray}

\end{theorem}

\begin{proof}
In order to characterize a sufficient scaling of feedback bits, we set the
rate gap upper bound given in Theorem \ref{thm-power-loss} equal to the maximum
allowable gap of $\log_2 b$:
\begin{eqnarray*}
\Delta R (P) < \log_2 \left( 1 + P \cdot 2^{-\frac{B}{M-1}} \right)
\triangleq \log_2 b.
\end{eqnarray*}
By inverting this expression and solving for $B$ as a function of 
$b$ and $P$ we get:
\begin{eqnarray}  \label{eq-N_FB}
B &=& (M-1) \log_2 P - (M-1)\log_2(b-1) \\
&=& \frac{ (M-1) \log_2 10}{10} P_{dB} -  (M-1)\log_2(b-1) \\
&\approx& \frac{M-1}{3} P_{dB} -  (M-1)\log_2(b-1). \label{eq-scaling}
\end{eqnarray}
With this scaling of feedback bits, we clearly have
$\Delta R (P) < \log_2 b$ for all $P$, as desired.
\end{proof}

The rate offset of $\log_2 b$ (per user)  can easily be translated into
a power offset, which is a more useful metric from
the design perspective.  Since a multiplexing gain of $M$
is achieved with zero-forcing, the zero-forcing curve has a
slope of $M$ bps/Hz/3 dB at asymptotically high SNR.
Therefore, a rate offset of $\log_2 b$
bps/Hz per user, or equivalently $M \log_2 b$ bps/Hz in throughput, 
corresponds to  a power offset of $3 \log_2 b$ dB
\cite{Shamai_Verdu}\cite{Lozano_Tulino_Verdu_High_SNR}.
Thus, $b=2$ corresponds to a 3 dB offset, and the
resulting scaling of bits takes on a particularly
simple form when a 3 dB offset is desired:
\begin{equation} \label{eq-3dB}
\setlength{\fboxsep}{.2cm}
 \fbox{$ \displaystyle
B = \frac{M-1}{3} P_{dB} ~~\textrm{bits/mobile}$.}
\end{equation}
In order to achieve a smaller power offset,
$b$ needs to be made appropriately smaller.  For example,
a 1-dB offset corresponds to $b=10^{1/10}=1.259$ and thus
an additional $1.95 (M-1)$ feedback bits are required at all SNR's.

\begin{figure}
\begin{center}
\centerline{\epsfig{file=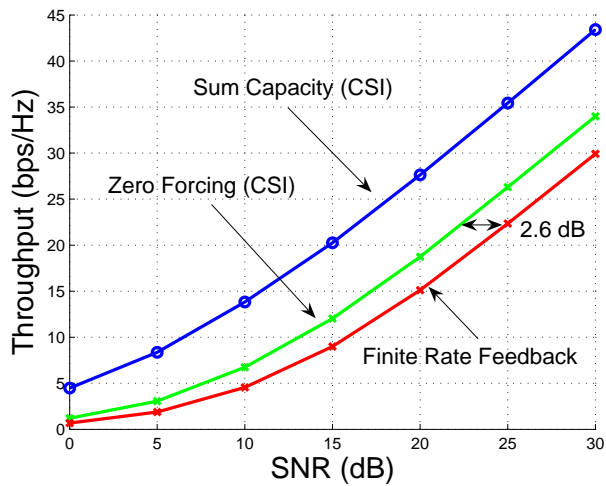, width=90mm}}
\end{center}
\caption{5 $\times$ 5 Channel with Increasing \# Feedback Bits}
\label{fig2}
\end{figure}

\begin{figure}
\begin{center}
\centerline{\epsfig{file=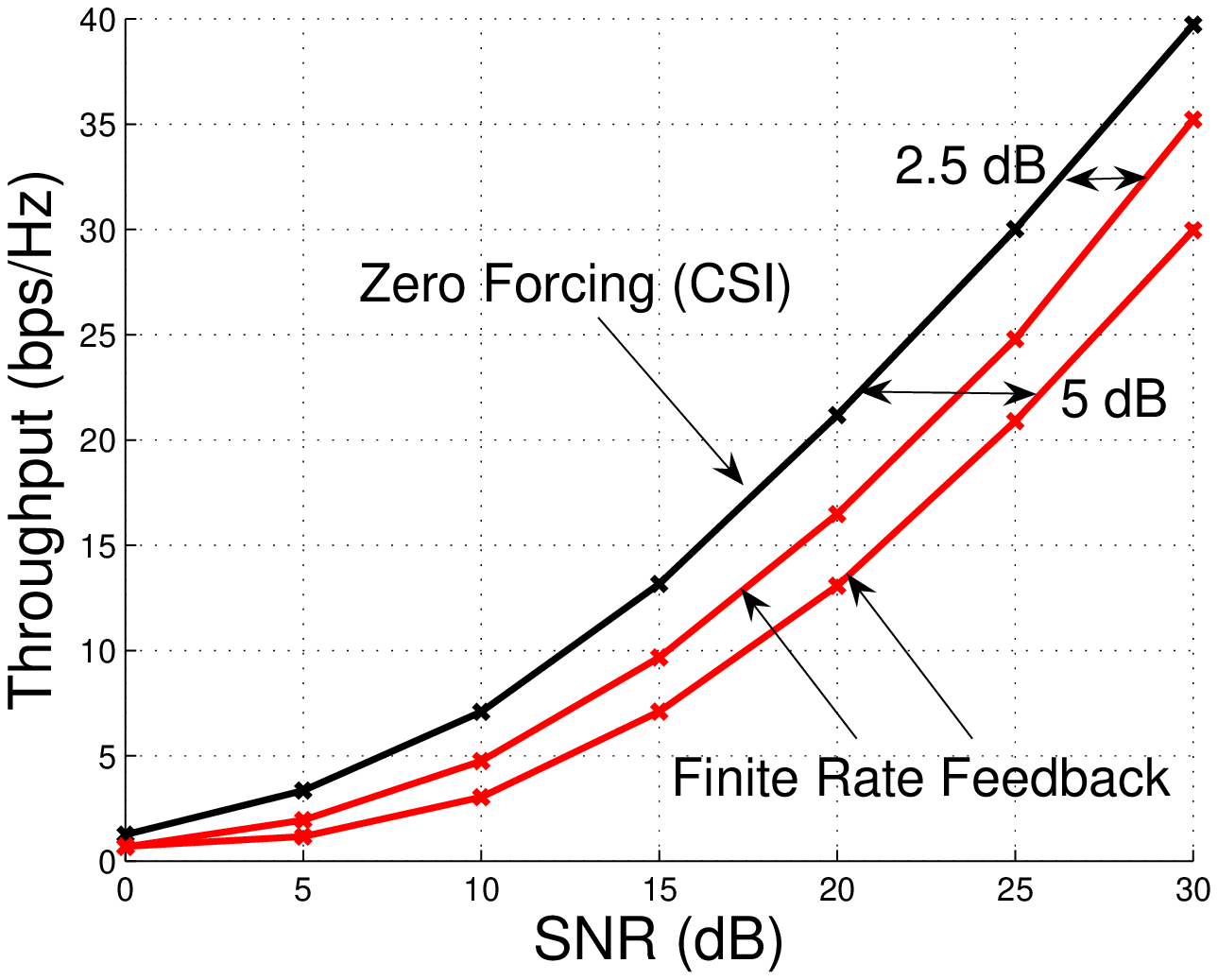, width=90mm}}
\end{center}
\caption{6 $\times$ 6 Channel with Increasing \# Feedback Bits}
\label{fig_6x6}
\end{figure}

In Fig. \ref{fig2}, throughput curves are shown
for a 5 antenna, 5 user system.  The feedback load
is assumed to scale according to the relationship given in (\ref{eq-3dB}),
and limited feedback is seen to perform within 2.6 dB of
perfect CSI zero-forcing.  Notice that the actual power offset
is smaller than 3 dB primarily due to the use of
Jensen's inequality in deriving the upper bound to $\Delta R(P)$
in Theorem \ref{thm-power-loss}.   The sum capacity, which
outperforms zero-forcing by 5.55 dB in a $5 \times 5$ system
\cite{Jindal_MIMO_BC_SNR}, is also shown. In Fig. \ref{fig_6x6}, throughput
curves are shown for a 6 antenna, 6 user system.  In this figure,
$B$ is scaled to guarantee a 3 dB and 6 dB gap from perfect
CSIT zero-forcing (i.e., $b = 2$ and $b=4$ in (\ref{eq-scaling_thm})).
Again, the actual gaps are smaller than the bounds (2.5 dB and 5 dB, 
respectively), but are still sufficiently close to make the bounds
useful.

If $B$ is scaled at a rate strictly greater than 
$(M-1) \log_2 P$, i.e. $B = \alpha \log_2 P$ for any $\alpha > M-1$,
the upper bound to the rate gap (Theorem \ref{thm-power-loss})
is easily shown to converge to zero:
\begin{eqnarray*}
\lim_{P \rightarrow \infty} \Delta R(P) \leq 
\lim_{P \rightarrow \infty}  \log_2 \left( 1 + P \cdot 2^{-\frac{B}{M-1}} \right) = 0,
\end{eqnarray*}
which implies that the rate gap itself converges to zero.
Thus, the throughput achieved with limited feedback converges (absolutely)
 to the perfect CSIT throughput at asymptotically high SNR.

However, scaling $B$ at a rate slower than $(M-1) \log_2 P$ results
in a strict reduction in the multiplexing gain, 
 as the following theorem shows:
\begin{theorem} \label{thm-mux_gain}
If $B$ is scaled as $B = \alpha \log_2 P$ for 
$\alpha < M-1$, the throughput curve achieves a multiplexing
gain of $M\left(\frac{\alpha}{M-1}\right)$.
\end{theorem}
\begin{proof}
See Appendix \ref{app-mux_gain}.
\end{proof}
Intuitively, the signal power grows linearly with $P$, while
the interference power scales as the product of $P$ and
the quantization error.  Since the quantization error is
of order $P^{- \frac{\alpha}{M-1}}$, the interference power scales as
$P^{\left(1-\frac{\alpha}{M-1} \right)}$, which gives an
SNR that scales as $P^{\frac{\alpha}{M-1}}$.  Thus, the resulting
multiplexing gain is  $M\left(\frac{\alpha}{M-1}\right)$.

\begin{figure}
\begin{center}
\centerline{\epsfig{file=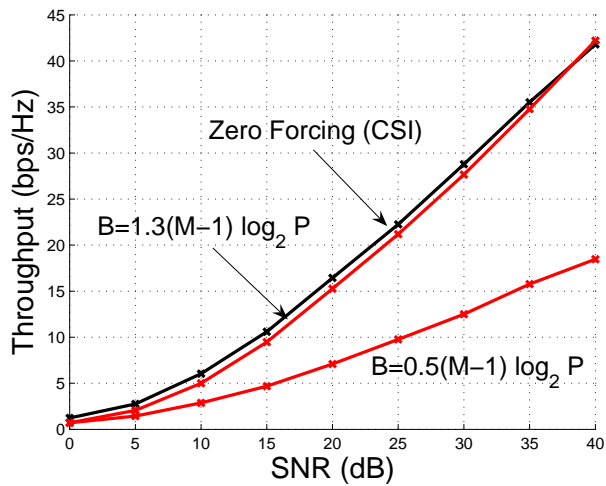, width=90mm}}
\end{center}
\caption{Multiplexing Gain as a Function of Feedback Scaling}
\label{fig-mux}
\end{figure}

In Figure \ref{fig-mux} the throughputs achieved with
feedback rates given by $B = 0.5(M-1) \log_2 P$
and $B = 1.3 (M-1) \log_2 P$ in a $4 \times 4$ channel
are shown.  When $\alpha = 0.5(M-1)$,
the achieved multiplexing gain is only $0.5 M = 2$, which corresponds
to a slope of 2 bps/Hz per 3 dB.  On the other hand, when 
$\alpha = 1.3(M-1)$, the full multiplexing gain is achieved and the 
rate gap relative to perfect CSIT zero-forcing converges to
zero at high SNR.

\subsection{RVQ vs. Optimal Vector Quantization} \label{sec-rvq_opt}

Given the previous sections results, an important issue to consider
is the possible performance loss incurred by use of random vector
quantization instead of an optimal vector quantizer.
In this section we derive an upper bound to the bit savings
that can result from using optimum vector quantization,
and find that the RVQ penalty is actually very small.
This is in accordance with results on
the optimality of RVQ for point-to-point MIMO and MISO
systems \cite{Santipach_Honig}.

In order to determine the sub-optimality of RVQ, we
utilize a lower bound to the quantization error of
\textit{any} vector quantization codebook developed
in \cite{Zhou_Wang_Giannakis, Mukkavilli_MIMO}:
\begin{lemma} \label{lemma-quant_bound}
Consider an arbitrary $B$-bit quantization codebook $\{ {\bf w}_1,\ldots,
{\bf w}_{2^{B}} \}$, and let the random variable $Y$ 
denote the quantization error
$Y = \sin^2(\angle(\tilde{\bh}_i, \hat{{\bh}_i}))$.  The random variable $Y$ 
\textit{stochastically dominates} the random variable $\tilde{Z}$,
whose CDF is given by:
\begin{eqnarray} \label{eq-cdf_lower}
F_{\tilde{Z}} (z) = \left \{ \begin{array}{ll} 2^{B} z^{M-1}, &
0 \leq z \leq 2^{- \frac{B}{M-1}} \\
1, & z \geq 2^{- \frac{B}{M-1}} \end{array} \right.
\end{eqnarray}
or  $F_{\tilde{Z}} (z) \geq F_Y (z)$ for all $0 \leq z \leq 1$.
\end{lemma}

This lemma applies to the quantization error for a \textit{fixed} 
quantization codebook, where the randomness is due only to the isotropic 
distribution of the channel vector. If RVQ is used, this result can be
applied to each quantization codebook, and therefore still holds
when considering the random variable describing the quantization error,
where there is randomness over both the channel realizations as well as the
codebooks:
\begin{corollary} \label{cor-quant_bound}
If the random variable $Z = \sin^2(\angle({\bh}_i, \hat{{\bh}_i}))$ 
denotes the quantization error achieved when using RVQ (as described in
Section \ref{sec-rvq}), then $Z$
stochastically dominates $\tilde{Z}$, whose CDF is given in (\ref{eq-cdf_lower}).
\end{corollary}

In order to determine the difference between RVQ and and a possibly
optimal quantizer,
we compute the feedback scaling required to maintain a bounded rate offset
as in Theorem \ref{thm-scaling}, but assuming the quantization error 
is described by $\tilde{Z}$ instead of $Z$.  In order to
compute this scaling, we first must compute the expectation of $\tilde{Z}$.
A simple calculation yields:
\begin{eqnarray*}
E[\tilde{Z}] &=& \int_0^1 \left( 1 - F_{\tilde{Z}}(z) \right) dz 
= \left(\frac{M-1}{M} \right) 2^{- \frac{B}{M-1}}.
\end{eqnarray*}
Notice that this differs from the upper bound to $E[Z]$ (Lemma \ref{lem-quant_upper})
only in the term $\frac{M-1}{M}$, and thus the interference
power is reduced by at most a factor of $\frac{M-1}{M}$.
In order to solve for the required feedback with this bound on
the quantization error we set:
\begin{eqnarray*}
\log_2 \left( 1 + P \cdot  \frac{M-1}{M} 2^{- \frac{B}{M-1}} \right) \triangleq
\log_2 b,
\end{eqnarray*}
and solve for $B$, which gives:
\begin{eqnarray}
B &=& (M-1) \log_2 P - (M-1)\log_2(b-1) -
(M-1) \log_2 \left( \frac{M}{M-1} \right).
\end{eqnarray}
Comparing this with the similar term for RVQ in (\ref{eq-N_FB}),
we see that the bit savings is a constant factor of
$(M-1) \log_2 \left( \frac{M}{M-1} \right)$ bits at all SNR's.
Furthermore, using the fact that $\log_e(1+x) \leq x$ for
$x > 0$ we have
$(M-1) \log_e \left( \frac{M}{M-1} \right) =
(M-1) \log_e \left( 1+ \frac{1}{M-1} \right) \leq
(M-1) \frac{1}{M-1} = 1$. Converting to base 2,
we see that
$(M-1) \log_2 \left( \frac{M}{M-1} \right) \leq \log_2 e$ bits,
or approximately 1.44 bits.  Thus, using RVQ leads to at most
a 1.44 bit penalty relative to optimum vector quantization, which
is quite small relative to the total feedback load.
We should note that this is somewhat of an approximation
because we have used Jensen's inequality to
derive the feedback load for both RVQ and for the lower
bound.  However, numerical results validate the accuracy
of Theorem \ref{thm-scaling}, and thus of this metric.


An alternative method to measure RVQ against optimum vector
quantization is to compare the performance for the same
number of feedback bits, as opposed to the above analysis in which
we compared the required feedback load required for identical performance.
As stated earlier, the quantization error, and thus the interference,
is a factor of $\frac{M-1}{M}$ smaller in the lower bound.
If the feedback is scaled in order to maintain a 3 dB
gap from perfect CSI zero forcing, the noise and interference
term are kept bounded by two (which corresponds to 3 dB).
The lower bound, on the other hand, would be $1 + \frac{M-1}{M}$
instead of 2.  For $M=5$, this corresponds to 2.55 dB, or
a 0.45 dB advantage relative to RVQ.  As the number of
transmit antennas $M$ increases, this gap clearly goes to zero.

\subsection{Regularized Zero-Forcing}

Although zero-forcing precoding performs quite well at moderate and high
SNR's, regularization can significantly increase throughput
 at low SNR's \cite{Peel2005}.  
In fact, this is exactly analogous to the difference between
zero-forcing equalization and MMSE equalization:
while zero-forcing results in complete cancellation of (inter-symbol)
interference, an MMSE equalizer alternatively allows a measured
amount of interference into the filtered output such that the
output SNR is maximized.  Regularized zero-forcing is also implemented
through a linear precoder, but with a slightly
different selection of beamforming vectors.  If $\hat{\bf H}$
denotes the concatenation of quantization vectors available to the
access point, the zero-forcing beamforming vectors are chosen as the
normalized columns of $\hat{\bf H}^{-1}$, or equivalently of
$\hat{\bf H}^{\dagger} (\hat{\bf H} \hat{\bf H}^{\dagger} )^{-1}$.
With regularized zero-forcing, the beamforming vectors ${\bf v}_1, \ldots, {\bf v}_K$ 
are chosen to be the normalized columns of the matrix:
\begin{eqnarray*}
\hat{\bf H}^{\dagger} \left(
\hat{\bf H} \hat{\bf H}^{\dagger} + \frac{M}{P} {\bf I} \right)^{-1}.
\end{eqnarray*}
The use of the regularization constant $\frac{M}{P}$ is well motivated
by results in \cite{Peel2005} as well as the optimal MMSE filters on the
dual multiple access channel \cite{Viswanath_Tse_Journal}.  It is clear
from this regularization that the regularized beamforming vectors
will converge to standard zero-forcing vectors at asymptotically
high SNR.

Since the rates achieved by zero-forcing and regularized zero-forcing
converge at asymptotically high SNR, the feedback scaling specified
in Theorem \ref{thm-scaling} gives the desired rate/power offset
for regualarized ZF at asymptotically high SNR.  Although we have not been
able to extend Theorems \ref{thm-power-loss} or \ref{thm-scaling}
to the rate offset between regularized ZF based on perfect versus
feedback-based CSIT, numerical results indicate that these results actually
hold at all SNR's, and not just at very high SNR values.  In fact,
numerical results indicate that Theorem \ref{thm-scaling} more 
accurately predicts the rate offset
for regularized ZF than for standard ZF at low and moderate SNR values.

\begin{figure}
\begin{center}
\centerline{\epsfig{file=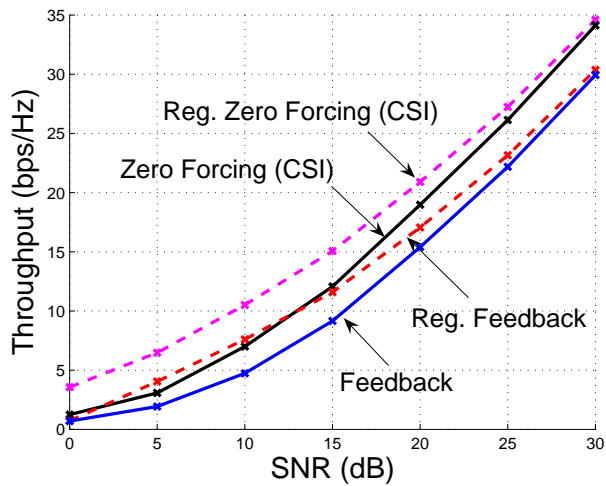, width=90mm}}
\end{center}
\caption{$5 \times 5$ Channel with Regularized Zero-Forcing}
\label{fig-reg55}
\end{figure}

Throughput curves for zero-forcing (solid lines) and regularized zero-forcing
(dotted lines) with perfect CSIT and feedback-based CSIT 
(with $B= \frac{M-1}{3} P_{dB}$) are shown
for a $5 \times 5$ channel in Figure \ref{fig-reg55}.
The throughput achieved with regularized ZF with perfect CSIT is significantly
larger (by approximately 4 bps/Hz) than the throughput with perfect CSIT based ZF 
for SNR's between
0 and 15 dB, but the two curves converge at very high SNR.  The same
is true for the throughput achieved with finite rate feedback and regularized
ZF vs. standard ZF.  Furthermore, while the rate offset between
perfect CSIT-ZF and feedback-CSIT ZF increases from nearly zero at low SNR
to its limiting value at high SNR, the rate offset between the two
regularized ZF curves is relatively constant over the entire SNR range, 
as noted earlier.


\section{Performance Comparison} \label{sec-comparison}

In this section we present numerical results comparing the throughput achieved
with finite rate feedback and two alternative transmission techniques for the
MIMO downlink, random beamforming and TDMA.  Because
regularized ZF outperforms ZF at all SNR's, we only consider regularized
ZF systems.   

Random beamforming is an extension of opportunistic beamforming
\cite{Pramod} to the multiple antenna downlink  \cite{Sharif_Hassibi}.  
The transmitter randomly chooses
$M$ orthogonal beamforming vectors and transmits pilot symbols along these
vectors.  Each mobile measures the SINR of each beamforming vector, and feeds
back the index of the vector with the highest SINR (requiring
$\log_2 M$ bits), along with the corresponding SINR.  The access point then
transmits to the best user on each of the beamforming vectors.
The required feedback per mobile is quite small ($\log_2 M$ bits plus
an analog SNR value, which will presumably be sufficiently quantized),
but this scheme does not perform well in systems with
a moderate number of mobiles (i.e., $K \approx M$).  In addition,
random beamforming is interference limited at asymptotically large SNR
if the number of mobiles is kept fixed.  

TDMA, in which the access point serves a single user at a time, is
perhaps the simplest downlink transmission scheme.  We consider the TDMA throughput 
achievable with perfect CSIT, in which the
access point transmits (using the capacity-achieving beamforming strategy)
to only the user with the largest SNR.
While it is possible to incorporate the effect of finite-rate
feedback into a TDMA system (as described in Section \ref{sec-miso}),
the effect is relatively negligible at the feedback levels considered
here and thus for simplicity we consider perfect CSIT.  Since a TDMA system
achieves a multiplexing gain of only one, we expect to see a significant
throughput degradation if TDMA is used, particularly at high SNR.
However, note that the difference between the sum capacity of the
MIMO downlink (achieved by DPC) and the achievable TDMA throughput
is not particularly large at SNR's less than 5 or 10 dB \cite{Jindal_Goldsmith_DPC}.

\begin{figure}
\begin{center}
\centerline{\epsfig{file=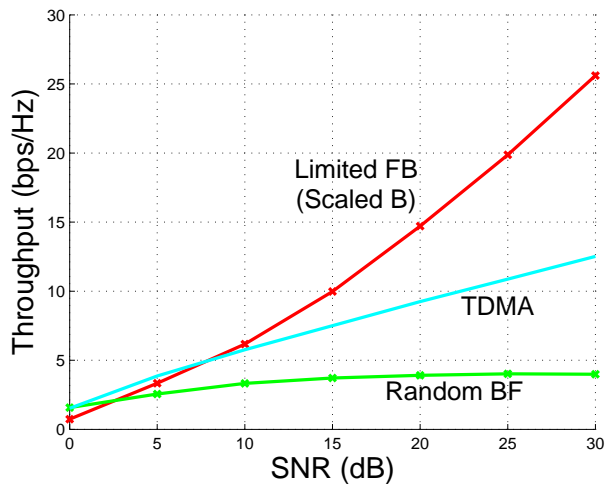, width=90mm}}
\end{center}
\caption{$4 \times 4$ Channel with Scaled Feedback, TDMA, \& Random BF}
\label{fig-compare44}
\end{figure} 

\begin{figure}
\begin{center}
\centerline{\epsfig{file=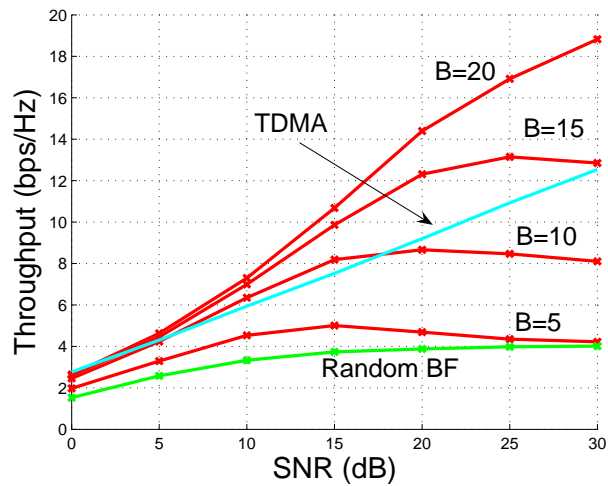, width=90mm}}
\end{center}
\caption{$4 \times 4$ Channel with Fixed Feedback, TDMA, \& Random BF}
\label{fig-compare44b}
\end{figure}

Figure \ref{fig-compare44} plots achievable throughput for
a finite-rate feedback system with $B$ scaled to maintain a 3 dB
offset (equation (\ref{eq-3dB})), TDMA, and random beamforming, in
a $4 \times 4$ channel.
Finite-rate feedback outperforms random BF beyond 5 dB, which is not surprising
given that the feedback level per mobile (which is given by $B = P_{dB}$ for this
particular channel) is significantly higher than for random BF.
Finite-rate feedback and TDMA give approximately the same throughput
up to 10 dB, after which the feedback system begins to significantly outperform
TDMA, due to the superior multiplexing gain of the zero-forcing system.
The same channel is considered in Figure \ref{fig-compare44b}, but with
fixed feedback levels ($B=5,10,15,20$) in the finite-rate feedback 
system\footnote{Note that the finite-rate feedback throughput decreases with
SNR in some cases due to the decreasing regularization factor as a function
of the SNR.  A more careful tuning of this parameter can prevent this
behavior, but does not significantly increase throughput.}.
Here we see that TDMA is a better choice than finite-rate feedback
with either 5 or 10 bits of feedback per mobile.  If 15 or 20 feedback
bits are permitted, however, finite-rate feedback can provide a significant
advantage over TDMA, particularly above 10 dB.

\begin{figure}
\begin{center}
\centerline{\epsfig{file=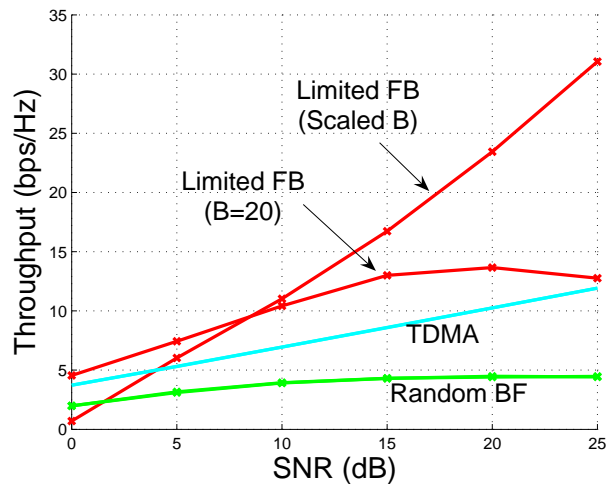, width=90mm}}
\end{center}
\caption{$8 \times 8$ Channel with Feedback, TDMA, \& Random BF}
\label{fig-compare88}
\end{figure} 

Figure \ref{fig-compare88} displays achievable throughput 
in a $8 \times 8$ channel for finite-rate
feedback systems with $B$ scaled according to equation (\ref{eq-3dB})
and with $B=20$, along with TDMA and random BF throughputs.
The throughput achieved with scaled feedback is considerably larger
than the TDMA throughput, due to the large
number of spatial degrees of freedom used by the zero-forcing system.
The 20 feedback bit system outperforms TDMA at moderate
SNR's, but hits the interference limited regime around 15 or 20 dB.

In general, finite-rate feedback based zero-forcing
outperforms TDMA at SNR's above 5 or 10 dB if the feedback per mobile
is sufficiently large, but does not provide a significant advantage 
at low SNR's.  Furthermore, random beamforming is outperformed by both
TDMA and finite-rate feedback at essentially all SNR levels.  However,
this is largely due to the limited number of mobiles, as random
beamforming does provide excellent performance in systems with many
more mobiles than AP antennas.

\section{Conclusions} \label{sec-conclusions}

The use of multi-user MIMO techniques can significantly
increase downlink throughput without requiring large numbers of
antennas at each mobile device.  However, it is crucial that the
transmitter have accurate channel state information in order to
realize these gains.  In fact, the availability of
channel state information appears to be the critical issue that
will determine the feasibility of multi-user MIMO techniques in
future wireless communication systems.

We investigated the finite-rate feedback
model, in which each mobile quantizes its channel realization to
a finite number of bits that are fed back to the transmitter.
Although this model has been extensively studied for point to
point MIMO channels, conclusions are quite different in the multi-user setting.
Our primary result showed that the number of feedback bits per mobile
must be increased linearly with the SNR (in dB) in order to
realize multi-user MIMO benefits.   As a result of this,
feedback levels are quite high in MIMO downlink channels:
in a 4 antenna system, for example, each mobile must feed
back 10 bits at an overall system SNR of 10 dB.   In contrast,
feedback does not need to scale with SNR in point to point
MIMO systems, and even a relatively small number of bits (e.g., 4 or 5)
can be very beneficial at all SNR levels.  
This scaling relationship is particularly 
troublesome because multi-user MIMO techniques are generally
most beneficial in the high SNR regime.

The intuition for the extreme sensitivity of the MIMO downlink channel
to imperfection in CSIT, as compared to point-to-point MIMO channels,
is actually quite straightforward.  In  point-to-point MIMO channels,
imperfect CSIT leads to mismatch between the input and the 
transmission modes of the channel and thus some ``wasted'' 
transmission power (e.g., power that is transmitted into the
nullspace of the channel), which reduces the received SNR but has
no other deleterious effect.
In a MIMO downlink channel, imperfect CSIT also
leads to mismatch between the input and the channel, but the effect of this
is increased multi-user interference, which is significantly more
harmful than a reduction in desired signal power.

There are, however, a number of reasons to be optimistic
regarding the issue of CSI feedback in downlink channels.  First note
that we have considered only the most basic iid Rayleigh block fading model,
and feedback rates can surely be decreased by exploiting spatial and
temporal correlation inherent in any physical fading process, as has
already been extensively studied for point to point MIMO channels
\cite{Love2004_Correlated}\cite{Xia_Giannakis}\cite{Raghavan_Sayeed}\cite{Banister_Zeidler}.
Furthermore, recent work has shown that a small number of mobile
antennas can be used to significantly improve quantization quality
and thereby decrease the required feedback rates 
\cite{Jindal_Limited_FB_Reduction}.  In addition, it may also be
possible to exploit multi-user diversity effects and thereby
reduce feedback rates in systems with a large number of mobiles
\cite{Sharif_Hassibi}\cite{Yoo_Jindal_Goldsmith}\cite{Swannack_Wornell}.
Within this body of work, one interesting idea is to have only a
subset of mobiles (e.g., mobiles meeting a fixed SNR threshold)
feed back their channel information 
\cite{Gesbert_Alouini}\cite{Sanayei_Norastinia}\cite{Swannack_Wornell}.
While this technique can significantly reduce the total number of 
feedback bits to be transmitted, it also requires
contention-based feedback (i.e., random access), 
which can lead to throughput reduction as well as additional latency on the
feedback channels.
Although we have considered only digital feedback methods, analog
feedback appears to have a number of attractive properties for the
downlink channel 
\cite{Marzetta_Hochwald_Training}\cite{Thomas_Feedback}\cite{Caire_Downlink}.
A careful comparison of analog and digital techniques remains to be
performed.


We close by mentioning a few important issues not considered in this work.
One key assumption made in this work is that the channel feedback
is \textit{instantaneous}.  Of course, there will be some non-zero delay associated
with transmitting feed back bits from the mobiles to the access point, and
this delay can be quite significant in fast fading (large Doppler spread) channels.
In fact, numerical results in \cite{Caire_Downlink} indicate that feedback delay 
can severely limit the performance of certain downlink transmission schemes
at even moderate levels of Doppler.  In addition, each mobile is assumed to have
perfect CSI, while there will be non-zero receiver estimation error in any
practical system.  This will clearly lead to additional imperfection in the
CSI provided to the access point, and could have significant effects.
Frequency-selective channels also require attention; consider related
work on frequency-selective point-to-point MIMO channels 
\cite{Choi_Heath}\cite{Khaled_Heath}.
Finally, note that we have only considered
frequency division duplexed systems.  Channel reciprocity can be
exploited to acquire downlink channel state information from the uplink
in time division duplexed (TDD) channels, although recent results
indicate that, somewhat counter-intuitively, TDD may in fact be less
 attractive than FDD from a channel state information perspective 
\cite{Marzetta_Hochwald_Training}.   Many of the tools used here appear to
be well suited to analyze the effect of imperfect CSIT in TDD systems
as well.

\section*{Acknowledgment}
The author would like to thank David Love for discussions and
for providing a preprint of reference \cite{Yeung_Love}.
In addition, the author acknowledges useful discussions with
Georgios Giannakis, Syed Ali Jafar, Taesang Yoo, and Giuseppe Caire
on limited feedback systems.


\appendices

\section{Expected Quantization Error} \label{app-quant_error}
In this appendix we provide an alternative proof for the closed-form
representation of the expected quantization error.  The following integral
representation for the beta function is given in \cite[pg. 5]{Gupta_Nadarajah}:
\begin{eqnarray*}
\beta \left( c, \frac{a}{b} \right) =
b \int_0^1 z^{a-1}\left( 1-z^b \right)^{c-1} dz, ~~ a>0,~b>0,~c>0.
\end{eqnarray*}
With $a=1$, $b=M-1$, and $c=2^{B}+1$ this yields:
\begin{eqnarray*}
\beta \left( 2^{B}+1, \frac{1}{M-1} \right) =
(M-1) \int_0^1 \left( 1-z^{M-1} \right)^{2^{B}} dz.
\end{eqnarray*}
Using the fact that 
$\textrm{Pr}\left(\sin^2 \left( \angle (\tilde{\bf h}_i, \hat{\bf h}_i) \right) \geq z\right) = (1-z^{M-1})^{2^{B}}$, we have:
\begin{eqnarray*}
E_{{\bf H}, {\mathcal W}} \left[
\sin^2 \left( \angle (\tilde{\bf h}_i, \hat{\bf h}_i) \right) \right]
&=& \int_0^1 (1-z^{M-1})^{2^{B}} dz \\
&=& \frac{1}{M-1} \beta \left( 2^{B}+1, \frac{1}{M-1} \right) \\
&=& \frac{ \frac{1}{M-1}\Gamma \left( 2^{B}+1\right) \Gamma \left( \frac{1}{M-1}\right)}{ \Gamma \left( 2^{B}+1 + \frac{1}{M-1} \right)} \\
&=& \frac{ 2^B \Gamma \left( 2^{B}\right) \Gamma \left(1+ \frac{1}{M-1}\right)}{ \Gamma \left( 2^{B}+1 + \frac{1}{M-1} \right)} \\
&=& 2^{B} \cdot \beta \left(2^{B}, \frac{M}{M-1}\right),
\end{eqnarray*}
where we have used the fundamental equality $\Gamma(z+1)=z \Gamma(z)$.

\section{Proof of Lemma \ref{lem-quant_upper}} \label{app-quant_upper}

The expected quantization error is given by \cite{Yeung_Love}:
\begin{eqnarray}
E_{{\bf H}, {\mathcal W}} \left[
\sin^2 \left( \angle (\tilde{\bf h}_i, \hat{\bf h}_i) \right) \right]
= 2^{B} \cdot \beta \left(2^{B}, \frac{M}{M-1}\right).
\end{eqnarray}
For $M=2$, we have
\begin{eqnarray*}
2^{B} \cdot \beta \left(2^{B}, 2 \right) &=&
 \frac{ 2^{B}  \Gamma(2^{B})\Gamma(2)}
{\Gamma \left( 2^{B} + 2\right)} \\
 &=& \frac{ 2^{B}  \Gamma(2^{B})}
{\Gamma \left( 2^{B} + 2\right)} \\
&=& \frac{ \left(2^{B}\right)! }{  \left(2^{B}+1\right)! } \\
&=& \left(2^{B}+1\right)^{-1} \\
&<& 2^{-B}.
\end{eqnarray*}

For $M > 2$:
\begin{eqnarray*}
2^{B} \cdot \beta \left(2^{B}, \frac{M}{M-1}\right) &=&
2^{B}  \frac{ \Gamma(2^{B}) \Gamma \left(1 + \frac{1}{M-1}\right)}
{\Gamma \left( 2^{B} + 1 + \frac{1}{M-1}\right)} \\
&\leq&  \frac{ 2^{B}  \Gamma(2^{B}) }
{\Gamma \left( 2^{B} + 1 + \frac{1}{M-1}\right)} \\
&=& \frac{ \Gamma(2^{B} + 1) }
{\Gamma \left( 2^{B} + 1 + \frac{1}{M-1}\right)}.
\end{eqnarray*}
The above inequality is reached because $\Gamma(x) \leq 1$ for $1 \leq x \leq 2$, due to the convexity of the gamma function \cite{Davis_Gamma}
 and the fact that $\Gamma(1)=\Gamma(2)=1$.
By applying Kershaw's inequality for the gamma function\cite{Kershaw}:
\begin{eqnarray*}
\frac{ \Gamma(x+s)}{\Gamma(x+1)} < \left(x + \frac{s}{2} \right)^{s-1} 
~~~ \forall x>0, ~ 0 < s < 1,
\end{eqnarray*}
with $x = 2^{B} + \frac{1}{M-1}$ and $s = 1 -  \frac{1}{M-1}$
we get
\begin{eqnarray*}
\frac{ \Gamma(2^{B} + 1) }
{\Gamma \left( 2^{B} + 1 + \frac{1}{M-1}\right)} &<& 
\left( 2^{B} + \frac{M}{2(M-1)} \right)^{-\frac{1}{M-1}}.
\end{eqnarray*}
Furthermore, the decreasing nature of the function 
$(\cdot)^{-\frac{1}{M-1}}$ gives a further upper bound of
$2^{-\frac{B}{M-1}}$.

\section{Proof of Lemma \ref{lem-exp_log}} \label{app-log}

Let $Z = \sin^2(\angle(\tilde{\bf h}_i, \hat{\bf h}_i))$ represent the quantization error.
As stated in Section \ref{sec-rvq},  $Z$ is the minimum of $2^{B}$ 
beta $(M-1,1)$ random variables with CCDF given by:
$\textrm{Pr}(Z \geq z) = (1 - z^{M-1})^L$,
where $L=2^{B}$.  We wish to compute $E[\log_e Z]$,
or equivalently $E[- \log_e Z]$.  Since $0 \leq Z \leq 1$, 
the random variable $- \log_e Z$ is non-negative with support $[0, \infty)$.  
Using the fact that $E[X] = \int_0^{\infty} \textrm{Pr}(X \geq x) dx$
for non-negative random variables and the binomial expansion, we have:
\begin{eqnarray*}
E[- \log_e Z] &=& \int_{0}^{\infty} \textrm{Pr}(-\log_e Z \geq z) dz \\
&=& \int_{0}^{\infty} \textrm{Pr}(Z \leq e^{-z}) dz \\
&=& \int_{0}^{\infty} 1 - (1 - e^{-z(M-1)})^L dz \\
&=& \int_{0}^{\infty} 1 - \sum_{k=0}^L {L \choose k} (-1)^k e^{-z(M-1)k} dz \\
&=& \int_{0}^{\infty} \sum_{k=1}^L {L \choose k} (-1)^{k+1} e^{-z(M-1)k} dz \\
&=&  \sum_{k=1}^L {L \choose k} (-1)^{k+1}  \left( \int_{0}^{\infty} e^{-z(M-1)k} dz \right) \\
&=&  \frac{1}{M-1} \sum_{k=1}^L {L \choose k}  \frac{(-1)^{k+1}}{k}    \\
&=&  \frac{1}{M-1} \sum_{k=1}^L  \frac{1}{k},
\end{eqnarray*}
where the final line follows from \cite[Section 0.155]{Ryzhik}.

Furthermore, since $\log_e a = \int_1^a \frac{1}{x} dx$, we have
$\log_e L \leq  \sum_{k=1}^L  \frac{1}{k} \leq \log_e L + 1$.
We multiply by $\log_2 e$ to translate to base 2, and thus get
$E[-\log_2 Z] = \frac{\log_2 e}{M-1} \sum_{k=1}^L  \frac{1}{k}$.  Using
$L=2^{B}$ we get the subsequent bounds.

\section{Proof of Theorem \ref{thm-mux_gain}} \label{app-mux_gain}
The multiplexing gain can be expanded as:
\begin{eqnarray} \nonumber
m &\triangleq& \lim_{P \rightarrow \infty} 
 \frac {M E_{{\bf H},W} [R_{FB}(P)] } {\log_2 P} \\
&=& \lim_{P \rightarrow \infty} \frac{M E_{{\bf H},W} \left[ \log_2 \left(1 + 
\frac{P}{M} | {\mathbf h}_i^{\dagger} {\mathbf v}_i |^2 +
\frac{P}{M} \sum_{j \ne i} | {\mathbf h}_i^{\dagger} {\mathbf v}_j |^2 \right) \right]}{\log_2 P} \nonumber \\&& \nonumber
- \lim_{P \rightarrow \infty} \frac{M E_{{\bf H},W} \left[ \log_2 \left(1 + 
\frac{P}{M} \sum_{j \ne i} | {\mathbf h}_i^{\dagger} {\mathbf v}_j |^2 \right) \right]}{\log_2 P} \nonumber \\
&=& M - \lim_{P \rightarrow \infty} \frac{M E_{{\bf H},W} \left[ \log_2 \left(1 + 
\frac{P}{M} \sum_{j \ne i} | {\mathbf h}_i^{\dagger} {\mathbf v}_j |^2 \right) \right]}{\log_2 P}. \label{eq-mux_gain}
\end{eqnarray}
The final step follows because
\begin{eqnarray*}
\log_2 \left(1 + \frac{P}{M} | {\mathbf h}_i^{\dagger} {\mathbf v}_i |^2 \right) 
\leq
\log_2 \left(1 + \frac{P}{M} | {\mathbf h}_i^{\dagger} {\mathbf v}_i |^2 +
\frac{P}{M} \sum_{j \ne i} | {\mathbf h}_i^{\dagger} {\mathbf v}_j |^2 \right) 
\leq \log_2 \left(1 + P || {\mathbf h}_i ||^2 \right),
\end{eqnarray*}
and the multiplexing gain of the upper and lower bounds 
are easily shown to be one.

We first show that $m \leq \frac{\alpha}{M-1}$ (using $L=2^{B} = P^{\alpha}$):
\begin{eqnarray*}
E_{{\bf H},W} \left[ \log_2 \left(1 + 
\frac{P}{M} \sum_{j \ne i} | {\mathbf h}_i^{\dagger} {\mathbf v}_j |^2 \right) \right] &\geq& E_{{\bf H},W} \left[ \log_2 \left(\frac{P}{M} | {\mathbf h}_i^{\dagger} {\mathbf v}_j |^2 \right) \right] \\
&=& \log_2 \left( \frac{P}{M} \right) + 
E_{\bf H}\left[ \log_2 ||{\bh}_i||^2 \right] + 
E_{{\bf H},W} \left[ \log_2 | \tilde{\mathbf h}_i^{\dagger} {\mathbf v}_j |^2 \right] \\
&=& \log_2 P + E_{\mathbf H, {\mathcal W}} \left[ \log_2 \left(
\sin^2 ( \angle(\tilde{\mathbf h}_i, \hat{{\bf h}}_i) ) \right) \right] + O(1) \\
&=&  - \log_2 P - \frac{\log_2 e}{M-1} \sum_{k=1}^L \frac{1}{k} + O(1)\\
&\geq& \log_2 P - \frac{1}{M-1} \log_2 L + O(1)\\
&=& \left(1  - \frac{\alpha}{M-1} \right) \log_2 P + O(1)\\
\end{eqnarray*}
where we have used Lemma \ref{lem-exp_log} to 
evaluate the expectation of the logarithm of the interference term.
Plugging this bound in to (\ref{eq-mux_gain}) gives 
 $m \leq M \left( \frac{\alpha}{M-1} \right)$.

To show $m \geq M \left( \frac{\alpha}{M-1} \right)$, we upper bound the limit
in (\ref{eq-mux_gain}) using the fact that 
$| \tilde{\mathbf h}_i^{\dagger} {\mathbf v}_j |^2 \leq \sin^2(\tilde{\bh}_i, \hat{\bh}_i)$ for any $i \ne j$ (from (\ref{eq-int_upper_bound}))
and Jensen's inequality:
\begin{eqnarray*}
E_{{\bf H},W} \left[ \log_2 \left(1 + 
\frac{P}{M} \sum_{j \ne i} ||{\bh}_i||^2 | \tilde{\mathbf h}_i^{\dagger} {\mathbf v}_j |^2 \right) \right] &\leq& 
E_{{\bf H},W} \left[ \log_2 \left(1 + 
\frac{P}{M} (M-1) ||{\bh}_i||^2 \sin^2(\tilde{\bh}_i, \hat{\bh}_i) \right) \right] \\
&\leq& \log_2 \left( 1 + P(M-1) E \left[\sin^2(\tilde{\bh}_i, \hat{\bh}_i) \right] \right) \\
&\leq& \log_2 \left( 1 + P(M-1) 2^{-\frac{B}{M-1}} \right) \\
&=& \log_2 \left( 1 + (M-1) P^{\left(1-\frac{\alpha}{M-1} \right)} \right) \\
\end{eqnarray*}
Thus we have
\begin{eqnarray*}
\lim_{P \rightarrow \infty} \frac{M E_{{\bf H},W} \left[ \log_2 \left(1 + 
\frac{P}{M} \sum_{j \ne i} | {\mathbf h}_i^{\dagger} {\mathbf v}_j |^2 \right) \right]}{\log_2 P} \leq M \left( 1 - \frac{\alpha}{M-1} \right),
\end{eqnarray*}
which gives $m \geq M \left( \frac{\alpha}{M-1} \right)$.

\bibliographystyle{IEEEtran}
\bibliography{capacity}

\begin{thebibliography}{10}
\providecommand{\url}[1]{#1}
\csname url@rmstyle\endcsname
\providecommand{\newblock}{\relax}
\providecommand{\bibinfo}[2]{#2}
\providecommand\BIBentrySTDinterwordspacing{\spaceskip=0pt\relax}
\providecommand\BIBentryALTinterwordstretchfactor{4}
\providecommand\BIBentryALTinterwordspacing{\spaceskip=\fontdimen2\font plus
\BIBentryALTinterwordstretchfactor\fontdimen3\font minus
  \fontdimen4\font\relax}
\providecommand\BIBforeignlanguage[2]{{%
\expandafter\ifx\csname l@#1\endcsname\relax
\typeout{** WARNING: IEEEtran.bst: No hyphenation pattern has been}%
\typeout{** loaded for the language `#1'. Using the pattern for}%
\typeout{** the default language instead.}%
\else
\language=\csname l@#1\endcsname
\fi
#2}}

\bibitem{Caire_Shamai}
G.~Caire and S.~Shamai, ``On the achievable throughput of a multiantenna
  {G}aussian broadcast channel,'' \emph{IEEE Trans. Inform. Theory}, vol.~49,
  no.~7, pp. 1691--1706, July 2003.

\bibitem{Jindal_Goldsmith_DPC}
N.~Jindal and A.~Goldsmith, ``Dirty paper coding vs. {TDMA} for {MIMO}
  broadcast channels,'' \emph{IEEE Trans. Inform. Theory}, vol.~51, no.~5, pp.
  1783--1794, May 2005.

\bibitem{Narula_Trott1}
A.~Narula, M.~J. Lopez, M.~D. Trott, and G.~W. Wornell, ``Efficient use of side
  information in multiple antenna data transmission over fading channels,''
  \emph{IEEE J. Select. Areas Commun.}, vol.~16, no.~8, Oct. 1998.

\bibitem{Love_Heath}
D.~Love, R.~Heath, and T.~Strohmer, ``Grassmannian beamforming for
  multiple-input multiple-output wireless systems,'' \emph{IEEE Trans. Inform.
  Theory}, vol.~49, no.~10, pp. 2735--2747, Oct. 2003.

\bibitem{Mukkavilli_MIMO}
K.~Mukkavilli, A.~Sabharwal, E.~Erkip, and B.~Aazhang, ``On beamforming with
  finite rate feedback in multiple-antenna systems,'' \emph{IEEE Trans. Inform.
  Theory}, vol.~49, no.~10, pp. 2562--2579, Oct. 2003.

\bibitem{Love_Heath_Santipach_Honig}
D.~Love, R.~Heath, W.~Santipach, and M.~Honig, ``What is the value of limited
  feedback for {MIMO} channels?'' \emph{IEEE Communications Magazine}, vol.~42,
  no.~10, pp. 54--59, Oct. 2004.

\bibitem{Sharif_Hassibi}
M.~Sharif and B.~Hassibi, ``On the capacity of {MIMO} broadcast channels with
  partial side information,'' \emph{IEEE Trans. Inform. Theory}, vol.~51,
  no.~2, pp. 506--522, Feb. 2005.

\bibitem{Yoo_Goldsmith}
T.~Yoo and A.~Goldsmith, ``On the optimality of multiantenna broadcast
  scheduling using zero-forcing beamforming,'' \emph{IEEE J. Select. Areas in
  Commun.}, vol.~24, no.~3, pp. 528--541, March 2006.

\bibitem{Swannack_Wornell}
C.~Swannack, E.~Uysal-Biyikoglu, and G.~Wornell, ``{MIMO} broadcast scheduling
  with limited channel state information,'' in \emph{Proceedings of Allerton
  Conf. on Commun., Control, and Comput.}, Oct 2005.

\bibitem{Santipach_Honig_CDMA}
W.~Santipach and M.~Honig, ``Signature optimization for {CDMA} with limited
  feedback,'' \emph{IEEE Trans. Inform. Theory}, vol.~51, no.~10, pp.
  3475--3492, Oct. 2005.

\bibitem{Santipach_Honig}
------, ``Asymptotic capacity of beamforming with limited feedback,'' in
  \emph{Proceedings of Int. Symp. Inform. Theory}, July 2004, p. 290.

\bibitem{Ding_Love_Zoltowski}
P.~Ding, D.~Love, and M.~Zoltowski, ``Multiple antenna broadcast channels with
  shape feedback and limited feedback,'' 2005, submitted to \emph{IEEE Trans.
  Sig. Proc.}

\bibitem{Yoo_Jindal_Goldsmith}
T.~Yoo, N.~Jindal, and A.~Goldsmith, ``Finite rate feedback {MIMO} broadcast
  channels with a large number of users,'' submitted to \textit{ISIT 2006}.

\bibitem{Weingarten_Steinberg_Shamai}
H.~Weingarten, Y.~Steinberg, and S.~Shamai, ``The capacity region of the
  {G}aussian {MIMO} broadcast channel,'' in \emph{Proceedings of Conference on
  Information Sciences and Systems}, March 2004.

\bibitem{mimo_bc_journal}
S.~Vishwanath, N.~Jindal, and A.~Goldsmith, ``Duality, achievable rates, and
  sum-rate capacity of {MIMO} broadcast channels,'' \emph{IEEE Trans. Inform.
  Theory}, vol.~49, no.~10, pp. 2658--2668, Oct. 2003.

\bibitem{Viswanath_Tse_Journal}
P.~Viswanath and D.~N. Tse, ``Sum capacity of the vector {Gaussian} broadcast
  channel and uplink-downlink duality,'' \emph{IEEE Trans. Inform. Theory},
  vol.~49, no.~8, pp. 1912--1921, Aug. 2003.

\bibitem{Yu_Cioffi}
W.~Yu and J.~M. Cioffi, ``Sum capacity of {Gaussian} vector broadcast
  channels,'' \emph{IEEE Trans. Inform. Theory}, vol.~50, no.~9, pp.
  1875--1892, Sept. 2004.

\bibitem{Costa}
M.~Costa, ``Writing on dirty paper,'' \emph{IEEE Trans. Inform. Theory},
  vol.~29, no.~3, pp. 439--441, May 1983.

\bibitem{Jindal_MIMO_BC_SNR}
N.~Jindal, ``A high {SNR} analysis of {MIMO} broadcast channels,'' in
  \emph{Proceedings of IEEE Int. Symp. Inform. Theory}, Sept 2005.

\bibitem{Cover_Broadcast}
T.~Cover, ``Broadcast channels,'' \emph{IEEE Trans. Inform. Theory}, vol.~18,
  no.~1, pp. 2--14, Jan. 1972.

\bibitem{Jafar_Goldsmith_Isotropic}
S.~Jafar and A.~Goldsmith, ``Isotropic fading vector broadcast channels: The
  scalar upperbound and loss in degrees of freedom,'' \emph{IEEE Trans. Inform.
  Theory}, vol.~51, no.~3, pp. 848--857, March 2005.

\bibitem{Lapidoth_Shamai}
A.~Lapidoth, S.~Shamai, and M.~Wigger, ``On the capacity of a {MIMO} fading
  broadcast channel with imperfect transmitter side-information,'' in
  \emph{Proceedings of Allerton Conf. on Commun., Control, and Comput.}, Sept.
  2005.

\bibitem{Zamir2002}
R.~Zamir, S.~Shamai, and U.~Erez, ``Nested linear/lattice codes for structured
  multiterminal binning,'' \emph{IEEE Trans. on Inform. Theory}, vol.~48,
  no.~6, pp. 1250--1276, 2002.

\bibitem{Erez2004}
S.~ten Brink and U.~Erez, ``A close-to-capacity dirty paper coding scheme,'' in
  \emph{Proceedings. Int. Symp. Inform. Theory}, 2004.

\bibitem{Philosof2003}
T.~Philosof, U.~Erez, and R.~Zamir, ``Combined shaping and precoding for
  interference cancellation at low {SNR},'' in \emph{Proceedings. Int. Symp.
  Inform. Theory}, 2003.

\bibitem{Yeung_Love}
C.~Au-Yeung and D.~J. Love, ``On the performance of random vector quantization
  limited feedback beamforming in a {MISO} system,'' to appear, \textit{IEEE
  Trans. Wireless Commun.}

\bibitem{Davis_Gamma}
J.~P. Davis, ``Leonhard {Euler's} integral: A historical profile of the gamma
  function,'' \emph{American Mathematics Monthly}, vol.~66, no.~10, pp.
  849--869, Dec. 1959.

\bibitem{Telatar}
E.~Telatar, ``Capacity of multi-antenna {Gaussian} channels,'' \emph{European
  Trans. on Telecomm. ETT}, vol.~10, no.~6, pp. 585--596, November 1999.

\bibitem{Jafar_Srinivasa}
S.~A. Jafar and S.~Srinivasa, ``On the optimality of beamforming with quantized
  feedback,'' submitted to \textit{IEEE Trans. Inform. Theory.}

\bibitem{Peel2005}
C.~Peel, B.~Hochwald, and A.~Swindlehurst, ``Vector-perturbation technique for
  near-capacity multiantenna multiuser communication-{P}art {I}: channel
  inversion and regularization,'' \emph{IEEE Trans. on Communications},
  vol.~53, no.~1, pp. 195--202, 2005.

\bibitem{Shamai_Verdu}
S.~Shamai and S.~Verdu, ``The impact of frequency-flat fading on the spectral
  efficiency of {CDMA},'' \emph{IEEE Trans. Inform. Theory}, vol.~47, no.~4,
  pp. 1302--1327, May 2001.

\bibitem{Lozano_Tulino_Verdu_High_SNR}
A.~Lozano, A.~Tulino, and S.~Verdu, ``High-snr power offset in multiantenna
  communication,'' \emph{IEEE Trans. Inform. Theory}, vol.~51, no.~2, pp.
  4134--4151, Dec. 2005.

\bibitem{Zhou_Wang_Giannakis}
S.~Zhou, Z.~Wang, and G.~Giannakis, ``Quantifying the power loss when transmit
  beamforming relies on finite rate feedback,'' \emph{IEEE Trans. on Wireless
  Commun.}, vol.~4, no.~4, pp. 1948--1957, 2005.

\bibitem{Pramod}
P.~Viswanath, D.~Tse, and R.~Laroia, ``Opportunistic beamforming using dumb
  antennas,'' \emph{IEEE Trans. Inform. Theory}, vol.~48, no.~6, pp.
  1277--1294, June 2002.

\bibitem{Love2004_Correlated}
D.~Love and R.~Heath, ``Grassmannian beamforming on correlated {MIMO}
  channels,'' in \emph{Proceedings of IEEE Globecom}, vol.~1, 2004, pp.
  106--110.

\bibitem{Xia_Giannakis}
P.~Xia and G.~B. Giannakis, ``Design and analysis of transmit-beamforming based
  on limited-rate feedback,'' in \emph{Proceedings of IEEE Vehicular Tech.
  Conf.}, 2004.

\bibitem{Raghavan_Sayeed}
V.~Raghavan, A.~Sayeed, and N.~Boston, ``Near-optimal codebook constructions
  for limited-feedback beamforming in correlated {MIMO} channels,'' submitted
  to \textit{ISIT 2006}.

\bibitem{Banister_Zeidler}
B.~Banister and J.~Zeidler, ``Feedback assisted stochastic gradient adaptation
  of multiantenna transmission,'' \emph{IEEE Trans. Wireless Commun.}, vol.~4,
  no.~3, pp. 1121--1135, May 2005.

\bibitem{Jindal_Limited_FB_Reduction}
N.~Jindal, ``A feedback reduction technique for {MIMO} broadcast channels,''
  submitted to \textit{ISIT 2006}.

\bibitem{Gesbert_Alouini}
D.~Gesbert and M.~S. Alouini, ``Selective multi-user diversity,'' in
  \emph{Proceedings of Int. Symp. on Signal Proc. and Inform. Technology}, Dec.
  2003.

\bibitem{Sanayei_Norastinia}
S.~Sanayei and A.~Nosratinia, ``Opportunistic downlink transmission with
  limited feedback,'' submitted to \textit{IEEE Trans. Inform. Theory}, Aug.
  2005.

\bibitem{Marzetta_Hochwald_Training}
T.~Marzetta and B.~Hochwald, ``Fast transfer of channel state information in
  wireless channels,'' submitted to \textit{IEEE Trans. Sig. Proc.} Preprint
  available at http://mars.bell-labs.com.

\bibitem{Thomas_Feedback}
T.~Thomas, K.~Baum, and P.~Sartori, ``Obtaining channel knowledge for
  closed-loop multi-stream broadband {MIMO-OFDM} communications using direct
  channel feedback,'' in \emph{Proceedings of IEEE Globecom}, 2005.

\bibitem{Caire_Downlink}
G.~Caire, ``{MIMO} downlink joint processing and scheduling: a survey of
  classical and recent results,'' in \emph{Proceedings of Workshop on
  Information Theory and its Applications, San Diego, CA}, 2006.

\bibitem{Choi_Heath}
J.~Choi and R.~W. Heath, ``Interpolation based transmit beamforming for
  {MIMO-OFDM} with limited feedback,'' \emph{IEEE Trans. on Signal Processing},
  vol.~53, no.~11, pp. 4125--4135, Nov. 2005.

\bibitem{Khaled_Heath}
N.~Khaled, B.~Mondal, R.~W. Heath, G.~Leus, and F.~Petre, ``Interpolation-
  based multi-mode precoding for {MIMO-OFDM} systems with limited feedback,''
  submitted to \textit{IEEE Trans. Wireless Commun.}

\bibitem{Gupta_Nadarajah}
A.~Gupta and S.~Nadarajah, \emph{Handbook of Beta Distribution and Its
  Application}.\hskip 1em plus 0.5em minus 0.4em\relax Marcel Dekker, Inc.,
  2004.

\bibitem{Kershaw}
D.~Kershaw, ``Some extensions of {W. Gautschi's} inequalities for the gamma
  function,'' \emph{Mathematics of Computation}, vol.~41, no. 164, pp.
  607--611, Oct. 1983.

\bibitem{Ryzhik}
I.~S. Gradshteyn, I.~M. Ryzhik, A.~Jeffrey, and D.~Zwillinger, \emph{Table of
  Integrals, Series, and Products, 6th Edition}.\hskip 1em plus 0.5em minus
  0.4em\relax Academic Press, 2000.

\end{thebibliography}

\end{document}